\documentclass[journal]{IEEEtran}
\usepackage{cite}

\ifCLASSINFOpdf
\usepackage[pdftex]{graphicx}
\graphicspath{{../../paper/pics/}}
\else
\fi
\usepackage{booktabs}
\usepackage{numprint}
\usepackage{multirow}
\usepackage{wasysym}

\usepackage{array}
\newcolumntype{P}[1]{>{\centering\arraybackslash}p{#1}}
\usepackage{longtable}

\usepackage{supertabular}
\usepackage{placeins} 

\usepackage{color,soul}

\usepackage{xcolor}
\usepackage{framed} 
\colorlet{shadecolor}{yellow!80}

\usepackage[hyphens]{url}
\usepackage[hidelinks]{hyperref}

\usepackage{acronym}
\newacro{cots}[COTS]{Commercial Off The Shelf}   
\newacro{iiot}[IIoT]{Industrial Internet of Things}   
\newacro{iot}[IoT]{Internet of Things}     
\newacro{it}[IT]{Information Technology}
\newacro{ot}[OT]{Operation Technology}
\newacro{ip}[IP]{Internet Procotol}  
\newacro{hmi}[HMI]{Human Machine Interface}
\newacro{plc}[PLC]{Programmable Logic Controller}  
\newacro{cnc}[C\&C]{Command \& Control}  
\newacro{ics}[ICS]{Industrial Control System} 
\newacro{ids}[IDS]{Intrusion Detection System}
\newacro{scada}[SCADA]{Supervisory Control And Data Acquisition}  
\newacro{swat}[\textit{SWaT}]{\textit{Secure Water Treatment}}  
\newacro{lstm}[LSTM]{Long Short-Term Memory}
\newacro{cps}[CPS]{Cyber-Physical System}
\newacro{wsn}[WSN]{Wireless Sensor Networks}
\newacro{arima}[ARIMA]{Auto-Regressive Integrated Moving Average}
\newacro{dos}[DoS]{Denial of Service}    
\newacro{uv}[UV]{Ultraviolet}    
\newacro{ro}[RO]{Reverse Osmosis}    
\newacro{svm}[SVM]{Support Vector Machine}    
\newacro{sssp}[SSSP]{Single Stage Single Point}
\newacro{ssmp}[SSMP]{Single Stage Multi Point}
\newacro{mssp}[MSSP]{Multi Stage Single Point}
\newacro{msmp}[MSMP]{Multi Stage Multi Point}
\newacro{mtu}[MTU]{Master Terminal Unit}
\newacro{uf}[UF]{Ultra Filtration}
\newacro{osint}[OSINT]{Open Source INTelligence}
\newacro{geoint}[GEOINT]{GEOgraphic INTelligence}
\newacro{pgp}[PGP]{Pretty Good Privacy}
\newacro{scratch}[SCRATCh]{SeCuRe and Agile Connected Things}
\newacro{cvss}[CVSS]{Common Vulnerability Scoring System Calculator}
\newacro{nist}[NIST]{National Institute of Standards and Technology}
\newacro{nvd}[NVD]{National Vulnerability Database}
\newacro{cve}[CVE]{Common Vulnerabilities and Exposures}
\newacro{coi}[CoI]{Corruption of Information}
\newacro{doi}[DoI]{Disclosure of Information}
\newacro{tos}[ToS	]{Theft of Service}
\newacro{rce}[RCE]{Remote Code Execution}
\newacro{pptp}[PPTP]{Point-to-Point Tunnelling Protocol}
\newacro{slip}[SLIP]{Serial Line Internet Protocol}
\newacro{dmz}[DMZ]{De-Militarised Zone}
\newacro{scada}[SCADA]{Supervisory Control And Data Acquisition}
\newacro{iam}[IAM]{Identity and Access Management}

\begin{document}
	%
	\title{The Global State of Security in Industrial Control Systems: An Empirical Analysis of Vulnerabilities around the World}
	%
	%
	%
	\author{Simon~Daniel~Duque~Anton,
		Daniel~Fraunholz,
		Daniel~Krohmer,
		Daniel~Reti,
		Daniel~Schneider,
		and~Hans~Dieter~Schotten 
		\thanks{This is a pre-print of a paper published in the \textit{IEEE Internet of Things Journal}.
Please cite as: \textbf{SD Duque Anton, D Fraunholz, D Krohmer, D Reti, D Schneider, and HD Schotten: \textit{The Global State of Security in Industrial Control Systems: An Empirical Analysis of Vulnerabilites around the World}, IEEE Internet of Things Journal, May 2021}}
	\thanks{S. D. Duque Anton was with the German Research Center for Artificial Intelligence. He is now with the comlet Verteilte Systeme GmbH and with the University of Kaiserslautern.}
	\thanks{D. Reti, D. Schneider and H. D. Schotten are with the German Research Center for Artificial Intelligence. H. D. Schotten is also with the University of Kaiserslautern.}
	\thanks{D. Krohmer was with the German Research Center for Artificial Intelligence. He is now with the Fraunhofer Institute for Experimental Software Engineering.}
	\thanks{D. Fraunholz was with the German Research Center for Artificial Intelligence and is with the University of Kaiserslautern.}}%
	
	%
	%

	\markboth{IEEE Internet of Things Journal, November 2020}%
	{Duque~Anton \MakeLowercase{\textit{et al.}}: The Global State of Security in Industrial Control Systems: An Empirical Analysis of Vulnerabilities Around the World}
	%



	\maketitle
	
	\begin{abstract}
		Operational Technology (OT)-networks and -devices,
		i.e. all components used in industrial environments,
		were not designed with security in mind.
		Efficiency and ease of use were the most important design characteristics.
		However,
		due to the digitisation of industry,
		an increasing number of devices and industrial networks is opened up to public networks.
		This is beneficial for administration and organisation of the industrial environments.
		However, 
		it also increases the attack surface,
		providing possible points of entry for an attacker.
		Originally,
		breaking into production networks meant to break an Information Technology (IT)-perimeter first,
		such as a public website,
		and then to move laterally to Industrial Control Systems (ICSs) to influence the production environment.
		However,
		many OT-devices are connected directly to the Internet,
		which drastically increases the threat of compromise,
		especially since OT-devices contain several vulnerabilities.
		In this work,
		the presence of OT-devices in the Internet is analysed from an attacker's perspective.
		Publicly available tools,
		such as the search engine \textit{Shodan} and vulnerability databases,
		are employed to find commonly used OT-devices and map vulnerabilities to them.
		These findings are grouped according to country of origin,
		manufacturer, and number as well as severity of vulnerability.
		More than \numprint{13000} devices were found,
		almost all contained at least one vulnerability.
		European and Northern American countries are by far the most affected ones.
	\end{abstract}
	

	%
	\IEEEpeerreviewmaketitle

	\section{Introduction}
	%
	%
	%
	%
	
	\label{sec:introduction}
	\IEEEPARstart{S}{tarting} in the 1970's,
	the term of \ac{scada} was coined to describe all control and monitoring in industrial networks,
	today also known as \ac{ot} networks.
	At first,
	\acp{ics} were created in an application-specific manner.
	Control in industrial environments was provided with fixed wiring and custom designs,
	since different enterprises had different control requirements.
	To reduce cost and effort required,
	while increasing the capabilities of \acp{ics},
	\ac{cots} devices were used.
	\acp{plc},
	small embedded computational devices controlling sensors and actuators of industrial machines,
	became well-established and made set-up in automation environments easier.
	Still,
	applications were highly use case- and operator-specific and \ac{ot} networks were physically separated from public networks,
	such as the Internet \cite{Igure.2006}.
	This separation provided a certain level of security,
	which is the reason why many industrial communication protocols,
	such as \textit{EtherCAT} \cite{ethercat.2018} and \textit{Modbus} \cite{MODICONInc..1996,  ModbusIDA.2006},
	did not provide encryption and authentication in their initial version.
	However,
	as the \ac{iot} converges into industrial applications,
	creating the \ac{iiot},
	the assumptions of application specificity and physical separation no longer hold true.
	The key enablers of \ac{iot} and \ac{iiot} alike are intercommunication and embedded computation,
	requiring networking capabilities.
	This in turn increased the attack surface of industrial environments \cite{Duque_Anton.2017a}.
	Even though several enterprises have to rely on legacy devices for organisational reasons,
	networking is becoming more open and interconnected.
	Consequently,
	networks and devices that were not designed with security in mind are exposed to public networks.
	The implications are severe,
	security has to be integrated into \ac{ot} as well.
	In general, most industrial organisations do not use \ac{iiot} solutions on a productive scale. Interconnectivity is a key enabler for anything \ac{iiot} and, in fact, a conditio sine qua non for \ac{iiot}. A shift towards a more interconnected, \ac{iiot}-based approach in industrial organisations with classic communication protocols transfers the security issues of those protocols into a more connected environment, consequently opening up attack vectors. 
	A common issue regarding the integration of the \ac{iiot} into industrial organisations lies in the gap between technologies that are available and the current and past state of the art.
	Currently,
	most industrial organisations rely on \ac{scada} protocols that were developed in the 1970's without any means for security,
	as discussed above.
	Changing such protocols in these expensive,
	application-specific and often difficult-to-access industrial environments has proven to be no easy task.
	Consequently,
	convergence from classic \ac{scada}-based control to the \ac{iiot} integrates existing protocols into novel solutions.
	For field level communication,
	\textit{Modbus/TCP} \cite{MODICONInc..1996,  ModbusIDA.2006} is a good fit.
	However,
	any interaction beyond the boundaries of a given shop floor should be controlled by gateway applications.
	Approaches to such gateways that are capable of translating established but insecure communication protocols into secure protocols are presented by industry~\cite{Dorofeev.2021} and science~\cite{Gundall.2021, duqueanton2019implementing} alike.
	As \textit{Gundall and Schotten} state,
	the life-cycle durations of industrial plants are designed for decades,
	thus,
	the impact of legacy devices has to be considered~\cite{Gundall.2021}.
	This holds for the associated protocols as well,
	as an update on the protocol suite for any given \ac{plc} is highly unlikely.
	It is expected that in the near future,
	\ac{iiot} environments will employ classic industrial communications protocols for control and monitoring,
	while algorithms,
	such as OPC-UA,
	that are designed with a strong security focus,
	are used for tasks such as inter-plant communication or collection and adaption of settings in a given environment.
	Consequently,
	a quick abandonment and replacement of legacy protocols seems unlikely.
	Instead,
	they have to be integrated into novel protocol environments.
	In order for \ac{iiot} solutions to work securely and safely,
	cyber security has to be in place.	
	In order to design countermeasures,
	a thorough understanding of the threat landscape and the severity of security issues is required.
	This work presents a field study analysing \ac{ics} devices exposed to the Internet and the known vulnerabilities they contain.
	The contribution of this work is in presenting a use case analysis from the perspective of an attacker.
	By means of \ac{osint},
	devices are enumerated and compared to public vulnerability databases.
	Any discovered vulnerability could be exploited as is by an attacker.
	The results of this analysis provide insight about the types of attacks commonly used that \acp{plc} are susceptible to,
	as well as an overview of the likelihood these \acp{plc} are connected to the Internet.
	This information can aid operators to assess the likelihood and type of attack the \ac{ot} environment could fall prey to,
	and aid in implementing counter measures.
	Furthermore,
	this analysis provides a methodology for operators to assess the attack surface of their \ac{ot} environments.
	Potential effects and severity of a successful attack can be derived by the metric developed and applied in the course of this work.
	This metric can be transported to any type of device in order to assess its susceptibility to attacks. \par
\begin{figure}[!ht]
		\centering
		\includegraphics[width=\linewidth]{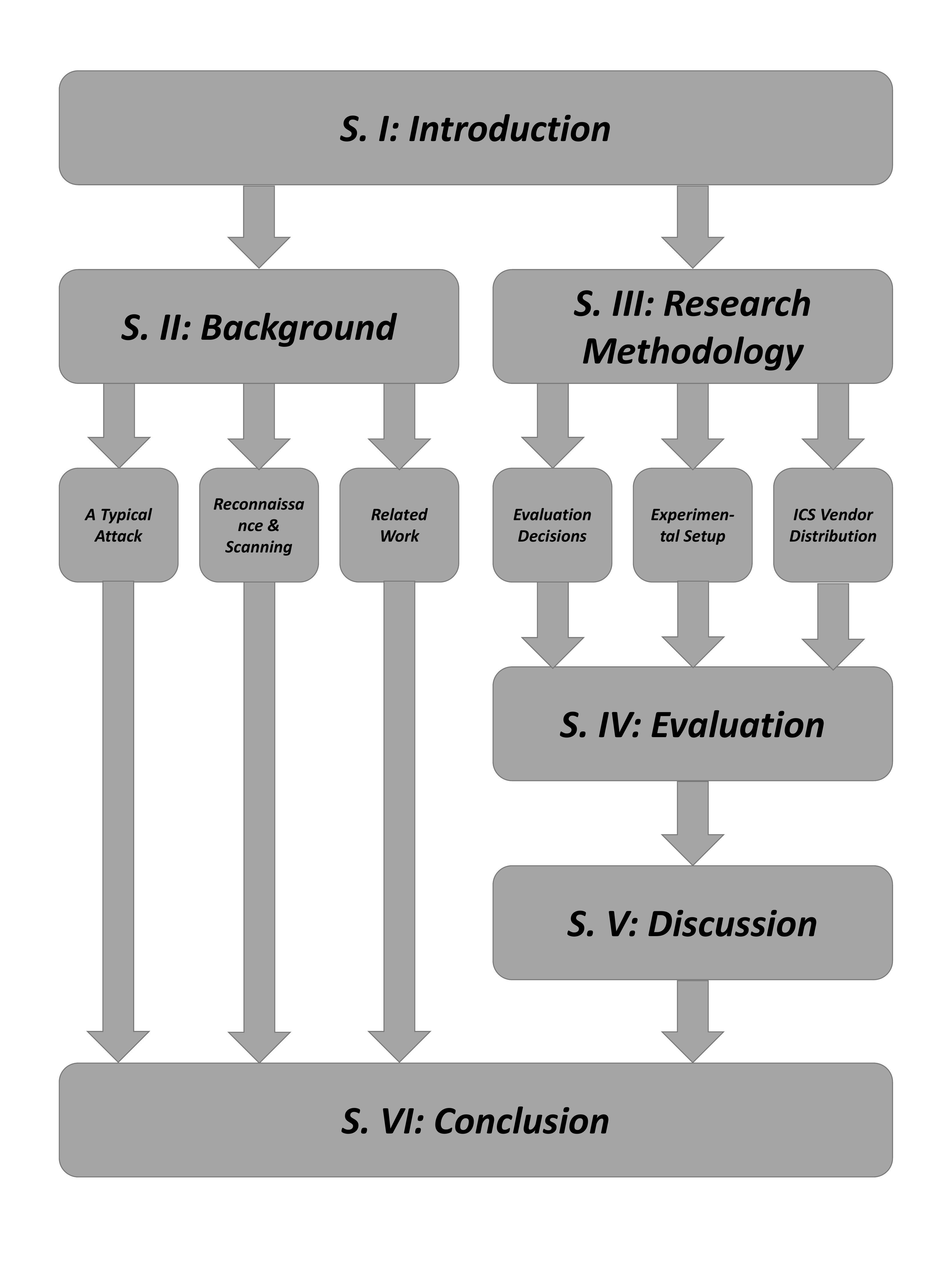}
		\caption{Structure of this Scientific Work}
		\label{fig:paper_structure}
	\end{figure}

 
	The contribution of this scientific tool is twofold:
	\begin{itemize}
		\item A thorough evaluation of \ac{ics} vulnerabilities based on actual experimental findings from an attacker's perspective. To the best of our knowledge, this has not been done before in this depth, although there has been a similar analysis of devices, limited to Japan~\cite{Abe.2016}.
		\item A methodology to systematically discover devices and assess their susceptibility to given \acp{cve}
		\item A quantitative evaluation and comparison of vulnerabilities in specific \ac{ot} devices, including severity and potential impact to derive a metric for  the threats to an organisation.
	\end{itemize}
	The remainder of this work is structured as follows.
	A background required for this work is discussed in  Section~\ref{sec:background},
	namely a  typical attack on industrial environments and methods of \ac{osint}.
	Section~\ref{sec:research_methodology} introduces the methodology to create the evaluation,
	which is presented in Section~\ref{sec:evaluation}.
	Section~\ref{sec:discussion} presents related work analysing the threat landscape of industrial environments,
	as well as an introduction and consideration of honeypots,
	and a discussion of the results.
	This work is concluded in Section~\ref{sec:conclusion}.
	A schematic overview of this structure is shown in Figure~\ref{fig:paper_structure}.

	\section{Background}
	\label{sec:background}
	Two concepts that are required for the context of this work are introduced in this section.
	First,
	a typical attack on industrial enterprises is presented.
	The characteristics of \ac{ot} networks lead to approaches of attackers which are different than in \ac{it}-based exploitation.
	Second,
	the concept of reconnaissance is discussed.
	This term is generally used in both security assessment and cyber attacks.
	Both attempt to find and exploit vulnerabilities in computer systems.
	A first step, according to the \textit{Lockheed Martin} cyber kill chain, \cite{Lockheed_Martin.2014} consists of reconnaissance which distinguishes between active and passive scanning.
	Reconnaissance is any activity aimed at gathering information about the target.
	After that, the distribution of the \ac{ics} market amongst prominent vendors is discussed as it provided the basis for selecting the devices for analysis.
	Finally, an overview of related works in the analysis of \ac{scada} and \ac{ot} security vulnerabilities is provided.

	\subsection{A Typical Attack on Industry}
\begin{figure}[!ht]
		\centering
		\includegraphics[width=\linewidth]{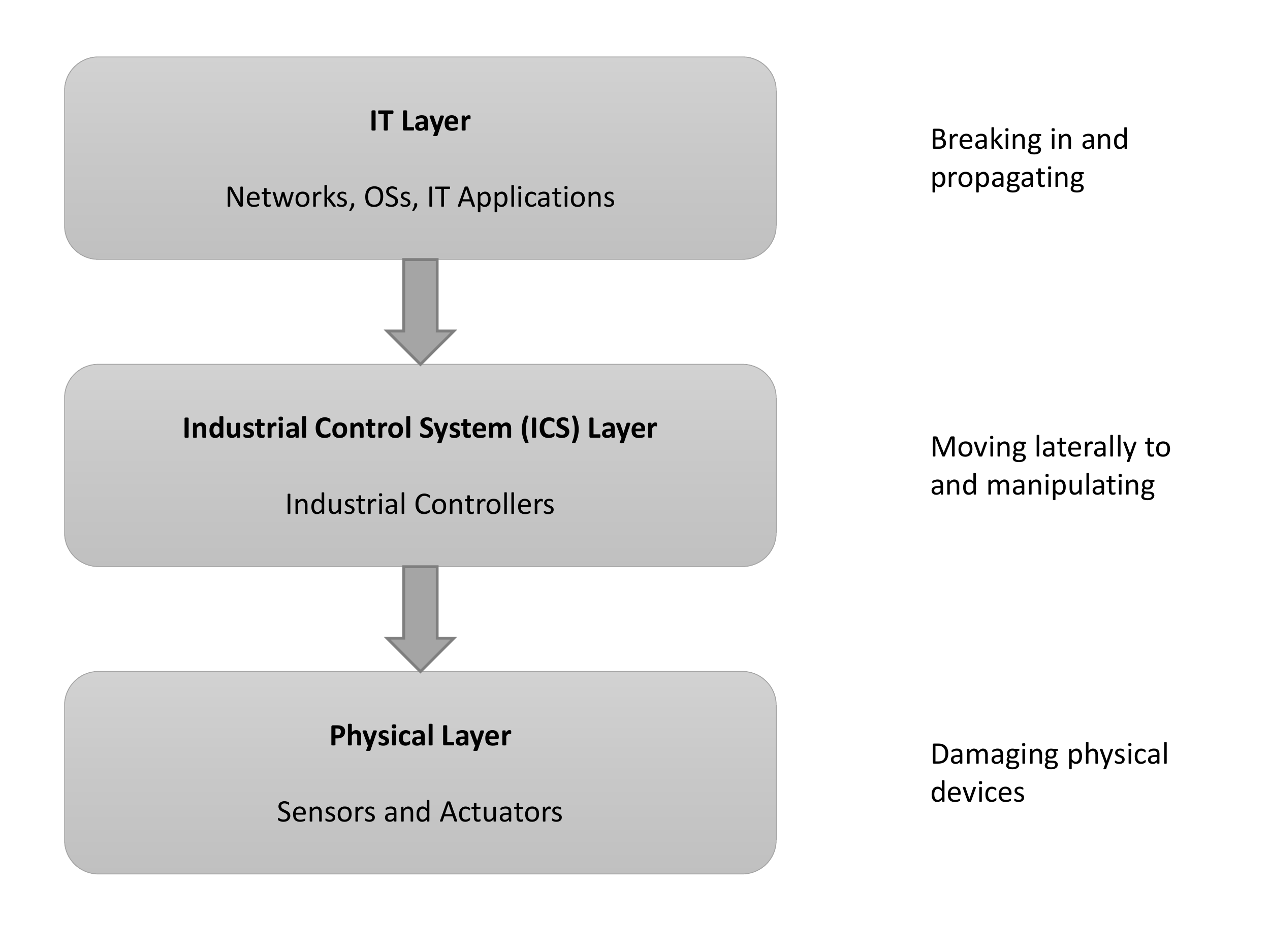}
		\caption{Stages of a Typical Attack on Industrial Applications}
		\label{fig:attack_stages}
	\end{figure}

	Typically,
	an industrial attack consists of three stages \cite{Langner.2013}.
	These stages are schematically depicted in Figure~\ref{fig:attack_stages}.
	At first,
	an \ac{it} layer has to be breached.
	This \ac{it} layer can constitute the public-facing website of an organisation or resources reachable from the intranet.
	Social engineering is the most common and most successful form of breaching the \ac{it} layer,
	e.g. with spear-phishing \cite{Parmar.2012}.
	Employees are tricked to open malware-infected resources by valid-looking e-mails or documents,
	thus enabling an attacker to infect their computer.
	Other,
	technology-based attack vectors on web-resources are collected and rated by the \textit{OWASP Foundation} \cite{OWASP.2020}. \par
	After an attacker has successfully breached this layer,
	traversal to the \ac{ics} plane is required,
	where control and monitoring of the \ac{ot} devices and their tasks occurs.
	Here,
	the attacker can exert influence on the \ac{ot} devices and tamper with the process description in acts of sabotage.
	Furthermore,
	theft of sensitive or protected information can occur at this place as well as acts of espionage and theft of intellectual property. \par
	The third level in this taxonomy is the physical layer.
	As the \ac{ot} devices in industrial applications are constituting \ac{cps},
	actions in the digital domain lead to actions in the physical domain.
	For example, controlling an automated drill will result in holes drilled in whichever material finds itself below the drill.
	This has strong implications on the security and safety of \ac{ot} devices,
	as this influence on the real world allows attackers to inflict property damage and potentially deadly injuries.
	Several well-known cyber attacks on \acp{ics},
	e.g. the \textit{Triton}-malware \cite{DiPinto.2018},
	the power outage in the Ukraine during December of 2015 caused by the malware \textit{BlackEnergy} \cite{Cherepanov.2016}, 
	and \textit{Stuxnet} \cite{Langner.2013},
	disrupted the physical process in order to achieve their goal.
	The fact that \ac{ot} devices and networks were not designed with security in mind \cite{Igure.2006} motivates the requirements to place \ac{ot} devices in internal networks.
	If access to public networks is given,
	the risk for cyber attacks is drastically increased.

	\subsection{Reconnaissance and Scanning}
	According to the \textit{Lockheed Martin} cyber kill chain \cite{Lockheed_Martin.2014},
	reconnaissance is the first stage of a cyber attack.
	It is performed to gain information about the intended target.
	The first step of reconnaissance,
	passive scanning,
	is a passive activity in the sense that no action is taken by the attacker to influence or interact with the system under investigation.
	Instead,
	public resources are employed and information is collected.
	Examples are methods such as \ac{geoint} and \ac{osint},
	for example whois domain enumeration,
	analysis of \textit{Google Maps} images,
	or mail and \ac{pgp} server queries. 
	Gathering information from social media and business networks is a means of \ac{osint}-based passive scanning as well.
	In summary,
	every activity aimed at gathering information about the intended target without the target having any chance to learn about it is called passive scanning.
	Penetration testers,
	i.e. hackers that test the security of networks,
	as well as criminals commonly put much effort into passive scanning as it is possible to obtain valuable information to be used in the exploitation without alerting the target.
	In the course of this work,
	\textit{Shodan} \cite{Shodan.2020} is used as a passive scanning tool.
	\textit{Shodan} is an Internet-based search engine that scans every device connected to the Internet and provides an interface to query this information.
	This feature explicitly extends to security- and privacy-relevant devices such as IP-cameras,
	\ac{iot} devices as well as \ac{ics} devices,
	just to name a few. \par
	Active scanning,
	in contrast to passive scanning,
	is performed with actively engaging target systems,
	such as host discovery and port scanning with network scanners like \textit{nmap} \cite{nmap.2020}.
	In active scanning, 
	an attacker actively queries the target systems in a fashion that the interaction is traceable.
	Therefore,
	active scanning is performed less extensively than passive scanning as it can warn a potential target about malicious activities.

	\subsection{Related Work}
	Assessment of vulnerabilities in \ac{ics} and \ac{ot} components from an attacker's point of view has not been performed in literature,
	to the best of the authors' knowledge.
	However,
	a variety of research addresses threats and risks for \acp{ics}.
	\textit{Gonzales et al.} performed an empirical analysis of security weaknesses in \acp{ics},
	based on disclosed vulnerabilities~\cite{Gonzales.2019}.
	This approach is somewhat similar to the approach performed in this work,
	however,
	the evaluation of how many of these vulnerabilities can actually be found in the wild is not performed by \textit{Gonzales et al.}.
	Results of this analysis seem to correlate with the findings of this work,
	however,
	the angle of analysis is different.
	\textit{Thomas and Clothia} categorise types of vulnerabilities based on historic \ac{scada} vulnerability data~\cite{Thomas.2020}.
	Based on this knowledge,
	recommendations on how to predict and protect against future vulnerabilities are proposed.
	In contrast,
	this work underlines the danger of well-known vulnerabilities.
	As long as old vulnerabilities are not fixed,
	they remain a threat to the organisation,
	and our research shows plenty of old vulnerabilities.
	\textit{Knowles et al.} evaluate known vulnerabilities,
	analyse the state of security in different industries and discuss methods and obstacles for introducing security measures in industrial environments~\cite{Knowles.2015}.
	Similarly, 
	\textit{Sullivan} analyses and evaluates attack vectors based on known attacks and introduces mitigation methods after elaborating differences between \ac{ics} and \ac{it} systems~\cite{Sullivan.2015}.
	A discussion on potential attacks on \acp{cps} from a theoretical control perspective is presented by \textit{Ding et al.}~\cite{Ding.2018}.
	\textit{Humayed et al.} present a survey of existing research on \ac{cps} security from different points of view:
	the security perspective,
	the \ac{cps} perspective,
	and the perspective of the system the \ac{cps} is embedded in~\cite{Humayed.2017}.
	The cybersecurity landscape of \acp{ics} is analysed by \textit{McLaughlin et al.}~\cite{McLaughlin.2016}.
	They evaluate the concepts for \ac{ics} security,
	consider past cyber attacks on \ac{ics},
	and provide an assessment for \ac{ics} security while looking at current test beds and trends in attack and defence methods.
	\textit{Holm at al.} put a focus on test beds for analysing attacks on and vulnerabilities of \acp{cps}~\cite{Holm.2015}.
	\textit{Abe et al.} evaluate the threats for \acp{ics} that are reachable from the internet, 
	based on observations in Japan~\cite{Abe.2016}.
	They explain how Internet-reachable \ac{ics}-devices can be found and exploited,
	without providing a quantitative overview and limited to Japan.
	Successful and attempted attacks on \acp{ics} are evaluated based on surveys by \textit{Luallen}~\cite{Luallen.2014}.
	Additionally,
	certain aspects of \ac{ics} and \ac{iiot} infrastructure are addressed in research.
	Testbeds,
	which can be used to research vulnerabilities,
	their exploitation and defences,
	are presented by \textit{Mathur and Tippenhauer}~\cite{Mathur.2016},
	\textit{Gardiner et al.}~\cite{Gardiner.2019},
	\textit{Gao et al.}~\cite{Gao.2013},
	and \textit{Hahn et al.}~\cite{Hahn.2010}.
	\textit{Skopik et al.} evaluate threats and vulnerabilities in smart metering,
	which brings \ac{ics} into the houses of users~\cite{Skopik.2012}.
	\textit{Plosz et al.} as well as \textit{Reaves and Morris} discuss vulnerabilities in wireless \ac{ics} communication by analysing the protocols for weaknesses and providing recommendations for securing the wireless communication~\cite{Plosz.2014,Reaves.2012}.
	Additionally,
	established security security research organisations regularly publish whitepaper containing statistics about \ac{ics} vulnerabilities,
	such as \textit{Kaspersky}~\cite{Kaspersky.2016},
	\textit{Positive Technologies}~\cite{PositiveTechnologies.2019},
	and the \textit{Control Systems Security Program} of the \textit{National Cyber Security Division}~\cite{HomelandSecurity.2010}.
	Such reports are based on statistical information obtained from organisations employing \ac{ics} equipment,
	or from studies of vulnerabilities.
	In contrast to this work,
	such studies rely on the participation of affected organisations and can only report incidents which have been disclosed.

	\section{Research Methodology}
	\label{sec:research_methodology}
	This section discusses the methodology that was applied to obtain,
	process, and evaluate the data.
	First,
	design decisions are presented that were applied in the course of this work.
	After that,
	the methodology to perform the individual case studies is presented.
	The selection of devices is motivated by the distribution of vendors in the \ac{plc} domain,
	which is presented in the third subsection.
	
	\subsection{Evaluation Decisions}
	In this section,
	several decisions made for the evaluation are introduced.
	These decisions influence the results of the analysis,
	while being necessary due to imperfections of the real world. \par
	Exposing PLC devices to the Internet is not recommended,
	it is expected that a significant number of \ac{ics} devices is not connected to the Internet.
	Thus,
	large amounts of false negative results are expected,
	i.e. devices that are operating in \acp{ics} environments,
	but are not found by \textit{Shodan}.
	This is expected as the given survey aims specifically at those \ac{ics} devices that are connected to the Internet.
	On the other hand,
	false positives are expected as well,
	for two reasons.
	First,
	some devices might be configured in a way that they do not use a TCP or UDP port in the standardised way.
	Every query,
	as discussed later on,
	relies on the port number in order to identify communication protocols.
	Non-standardised usage of said ports can introduce false positive as well as false negative results.
	However,
	each query has additional aspects and the results are further analysed.
	Ports are used as an initial indicator for potentially vulnerable devices.
	After they are discovered,
	additional information,
	such as server banners with descriptions of the running services and versions,
	are employed to ensure the correct analysis of the device.
	Also,
	further insight from other ports on the same device are taken into account,
	for example web services with an HTTP banner.
	Thus,
	false positives due to non-standardised usage of ports are expected to not have a significant influence on the result.
	Second,
	some organisations employ honeypots,
	e.g. Conpot,
	that are capable of mimicking \ac{ics} protocols.
	A discussion of honeypots in \ac{ics} devices is presented in Section~\ref{sec:discussion}. \par
	Furthermore,
	many vulnerabilities correspond to specific firmware versions,
	while later versions patch the given vulnerability.
	Some vendors might patch old firmwares as well without changing the version number,
	making these vulnerabilities obsolete.
	This might lead to false positive results.
	However,
	we have found no evidence of such practice during this research.
	Additionally,
	there are cases where web services are vulnerable and the security advisory of the vendor advises network segmentation,
	firewalling and deactivation of services.
	These services are considered vulnerable in the course of this work.
	Furthermore,
	the results are presented in tables which contain the \acp{cve} as well as the conditions required to be vulnerable and an indicator of how well the vulnerability to a \ac{cve} can be derived.
	The indicator is a circle with the following meaning:
	\begin{itemize}
		\item \CIRCLE: yes - The susceptibility to a vulnerability can be derived with certainty on measurable features
		\item \LEFTcircle: partially - The susceptibility to a vulnerability can be guessed based on sound assumptions
		\item \Circle: no - The susceptibility to a vulnerability cannot be derived with certainty based on the available information
	\end{itemize}
	
	\subsection{Experimental Setup}
	\label{sec:experimental_setup}
\begin{figure}[!ht]
		\centering
		\includegraphics[width=0.6\linewidth]{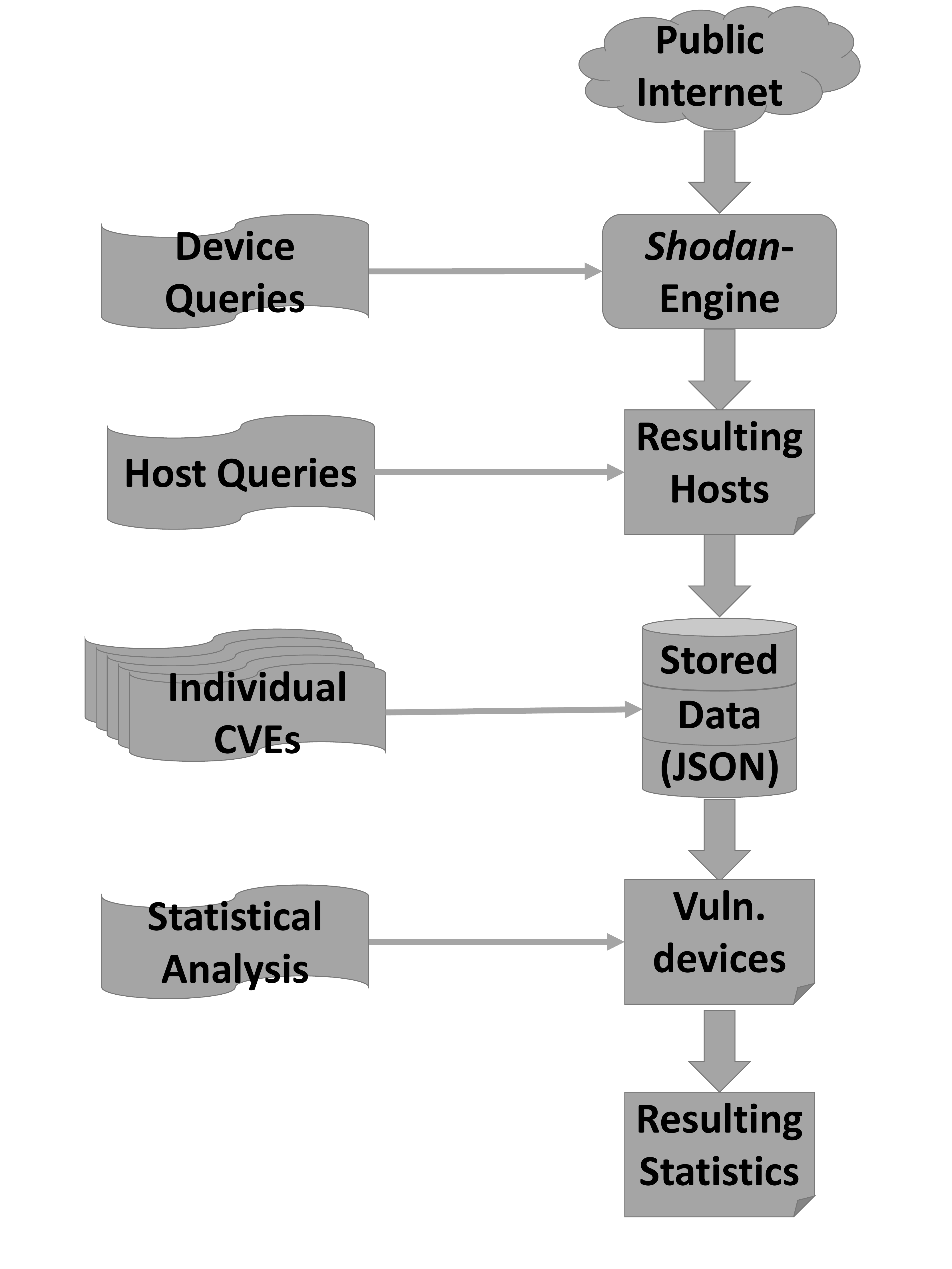}
		\caption{Schematic Overview of Evaluation Process}
		\label{fig:approach_overview}
	\end{figure}

	The selected devices,
	which were presented in Section~\ref{sec:research_methodology},
	were queried with the Search Engine \textit{Shodan} \cite{Shodan.2020} in an iterative fashion.
	A schematic overview of the process is shown in Figure~\ref{fig:approach_overview}.
	First,
	the query to search for was designed by manual inspection and interaction.
	In this work,
	five \ac{plc} series from different vendors were analysed.
	The corresponding queries were meant to catch any device that fits the respective series under investigation,
	based on banner and open ports.
	For each resulting host found with a query,
	this host was scanned with \textit{Shodan} via the \textit{Python}-API in order to obtain an overview of open ports and provided services and banners.	
	The results were stored in a JSON-formatted file,
	for which the \textit{json} library of \textit{Python} was used.
	In a second step,
	the conditions for each \ac{cve} were manually derived from several sources,
	mainly the \ac{nist} \ac{nvd} \cite{NIST.2020},
	but also security advisories of the vendors.
	These conditions often include firmware versions which could be derived from the banner,
	open ports or services or certain modules present on the host,
	also derived from banner information.
	In a final step,
	the prerequisites for each \ac{cve} that is potentially applicable to a given device are obtained from the JSON-formatted information by an automated \textit{Python}-script.
	The results are then used to calculate metrics,
	such as the amount of devices and vulnerabilities per country.
	For this,
	a custom made \textit{Python} script was employed.
	In order to make the process clear,
	it is described in the following toy example for the \textit{PerfectToast}-series of the fictional vendor \textit{KitchenAppliances}.
	Research showed that the \textit{PerfectToast}-series always runs a service at port 4567 and also displays a banner that says "perfect toast",
	so the query ``port:4567 perfect toast'' would be used.
	This would detect all potential devices of the \textit{PerfectToast}-series.
	The hosts are stored in a csv-type file and in a next step queried for any information that can be obtained of the given host,
	e.g. additional open ports and provided services.
	This would obtain the information for every \textit{PerfectToast}-device,
	such as open ports and services banners.
	Let's assume \textit{PerfectToast}-device have an open HTTP-port with number 80 and an open SSH-port with number 22 and return a web server version, 
	as well as an SSH-version upon query.
	All of these devices are then stored with the given host information in a JSON-file.
	After that,
	all \acp{cve} for which \textit{PerfectToast}-devices are potentially vulnerable,
	are collected manually,
	together with conditions related to the vulnerability.
	For example, the \textit{PerfectToast}-devices are vulnerable to a \ac{cve} if the Apache server version on HTTP port 80 is less than 2.9.
	Unfortunately,
	\acp{cve} are structured in a way that it is difficult to automatically discover any conditions that are explained in non-structured text.	
	In a final step,
	the devices stored in the JSON-file are queried to evaluate if they match the conditions for the \acp{cve}.
	So each of the \textit{PerfectToast}-devices which was found by \textit{Shodan} is queried for having an open HTTP port and an Apache server version of less than 2.9.
	If this condition is met,
	the device is considered to be vulnerable.
	In this fashion,
	each device is queried for every potential \ac{cve}. Subsequently the results from those queries undergo statistical analysis, thus providing a valuable insight into the factual state of vulnerability of \acp{ics}.	
	Furthermore,
	a categorisation of the results into six classes was performed,
	based on the aim of a successful attack.
	For this,
	the well-established STRIDE-model~\cite{Kohnfelder.1999} has been applied.
	STRIDE is an anagram describing the different types of computer security threats:
	\begin{itemize}
		\item \textbf{S}poofing, indicated in tables by S
		\item \textbf{T}ampering, indicated in tables by T
		\item \textbf{R}epudiation, indicated in tables by R
		\item \textbf{I}nformatin disclosure, indicated in tables by I
		\item \textbf{D}enial of service, indicated in tables by D
		\item \textbf{E}levation of privilege, indicated in tables by E
	\end{itemize}
	In general,
	an attacker can tamper,
	i.e. break the integrity of data to influence the process,
	disclose,
	i.e. steal information,
	disrupt the service,
	or elevate privileges to increase the radius and impact of operation.
	Of course,
	there might be an overlap,
	e.g. \ac{dos} by changing process parameters fits D as well as T,
	however,
	since the intended result is of interest in this context,
	this instance would be labelled as D.
	Spoofing at this point is of little interest,
	as many applications do not have identity management at all.
	Furthermore, 
	Repudiation is seldomly the goal of an attack,
	but rather a means to an end.
	All results obtained in the course of this work are presented in Section~\ref{sec:evaluation}.

	\subsection{Distribution of Vendors for ICS devices}
\begin{figure}[!ht]
		\centering
		\includegraphics[width=\linewidth]{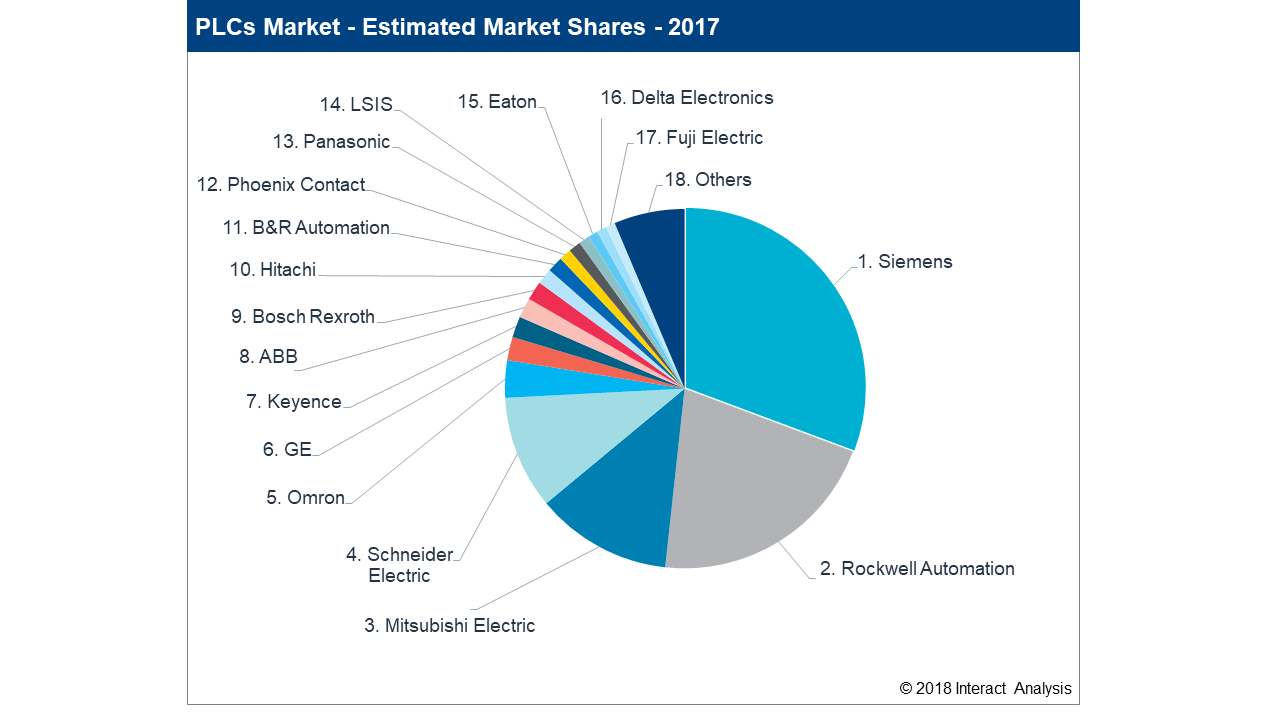}
		\caption{Estimated Market Distribution of PLC vendors according to \textit{Dawson}~\cite{Dawson.2017}}
		\label{fig:vendor_distribution}
	\end{figure}

	The market for \ac{ics} devices is mostly covered by five vendors.
	For the year 2017,
	the distribution is shown in Figure~\ref{fig:vendor_distribution} which was created by \textit{Dawson}~\cite{Dawson.2017}.
	The market shares of \ac{plc} vendors based on information from \textit{statista} is shown in Table~\ref{tab:market_shares_statista}~\cite{statista.2017}.
	\begin{table}[h!]
		\renewcommand{\arraystretch}{1.3}
		\caption{Market Shares of \ac{plc} vendors in 2017 according to \textit{statista}~\cite{statista.2017}}
		\label{tab:market_shares_statista}
		\centering
		\scriptsize
		\begin{tabular}{l r} 
			\toprule
			\textbf{Vendor} & \textbf{Market Share} \\
			\textit{Siemens} & 31\% \\
			\textit{Rockwell Automation} & 22\% \\
			\textit{Mitsubishi Electric} & 14\% \\
			\textit{Schneider Electric} & 8\% \\
			\textit{Omron} & 6\% \\
			\textit{B\&R Industrial Automation} & 4\% \\
			\textit{GE} & 3\% \\
			\textit{ABB Ltd.} & 2\% \\
			\textit{Others} & 10\% \\
			\bottomrule
		\end{tabular}
	\end{table}
	This matches the distribution shown in Figure~\ref{fig:vendor_distribution}.
	Furthermore,
	commercial market analysis,
	such as \textit{Mordor Intelligence} and \textit{PR Newswire} name 
	\textit{ABB Ltd.}, 
	\textit{Mitsubishi Electric}, 
	\textit{Schneider Electric}, 
	\textit{Rockwell Automation}, 
	and \textit{Siemens}~\cite{MordorIntelligence.2020}, or
	\textit{Schneider Electric}, 
	\textit{Rockwell Automaton}, 
	\textit{Siemens}, 
	\textit{Mitsubishi Electric}, and
	\textit{Omron}~\cite{Cision.2020} respectively as the most prominent vendors of \acp{plc}.
	Thus,
	we chose to analyse products or product groups from the following five vendors,
	as they represent a large share of the \ac{plc} market:
	\begin{itemize}
		\item \textit{Siemens}
		\item \textit{Rockwell Automation}
		\item \textit{Rockwell Automation}
		\item \textit{Mitsubishi}
		\item \textit{Schneider Electric}
		\item \textit{Omron}
	\end{itemize}
		This aims at addressing devices with a high distribution and application in industry.
	Since each group or type of devices from a specific vendor contains different potential vulnerabilities,
	exemplary devices or groups of devices have been selected for the study. 
	These devices should be easily identifiably with \textit{Shodan},
	have vulnerabilities that can be checked against,
	and be available in a number large enough to be expressive.
	Pre-emptive analysis of device types by the individual vendors led to the devices that are analysed below.

	\section{Evaluation}
	\label{sec:evaluation}
	This section evaluates the devices presented according to the methods discussed in Section~\ref{sec:research_methodology}.
	An overview of the state of security with respect to these devices is obtained. 
	
	\subsection{\textit{Schneider Electric BMX P34 series}}
	\label{ssec:schneider_electric}
	One of the devices evaluated was the Schneider Electric BMX P34 series \cite{SE.2020}.
	Schneider Electric,
	who,
	under the name Modicon developed the Modbus-protocol \cite{MODICONInc..1996},
	is the fourth largest manufacturer of \ac{ics} technology,
	as indicated by Figure~\ref{fig:vendor_distribution}.
	It was founded in 1836 and has its main seat in Paris.
	Schneider Electric is active in about 150 countries and thus has a significant influence on the industrial landscape.
	The Schneider Electric BMX P34 series is part of the \textit{Modicon M340} series,
	containing seven \ac{plc} devices with CPUs and different connectivity options. 
	\par 
	The devices were fingerprinted by \textit{Shodan} by looking for two conditions:
	First,
	the device had to provide Modbus-functionality.
	This was analysed by the open ports,
	standard port for Modbus/TCP is 502.
	If a service is active on that port,
	it is assumed to communicate with the Modbus-protocol.
	Furthermore,
	each device should have the string ``Schneider Electric BMX'' in their banner.
	The ``Schneider Electric BMX'' are part of the Modicon M340-series by Schneider Electric.
	If these two conditions were met,
	the device was employed for the analysis.
	The resulting search term was:
	\textit{port:502 ``Schneider Electric BMX''}.
	Since all banners were expressive,
	it is safe to assume that every device found was actually communicating via Modbus on port 502.
	\par 
	In general,
	devices of this series can contain a total of 59 different vulnerabilities,
	listed as \acp{cve} according to the \ac{nist} \ac{nvd} \cite{NIST.2020}.
	These vulnerabilities are listed, 
	with \ac{cvss} score version 3.1 and 2 taken from the respective vulnerability description \cite{NIST.2020} in Table~\ref{tab:methodology_vulnerabilities_schneider_electric_overview}.
	If a \ac{cve} did not have a \ac{cvss} score version 3.1,
	it was calculated manually and indicated with a star.  
	\vspace{1em}

	\renewcommand{\arraystretch}{1.3}
	%
	\tablefirsthead{%
		\toprule
		\textbf{Cl.} & \textbf{CVE} & \multicolumn{2}{r}{\textbf{CVSS}} & \multicolumn{2}{c}{\textbf{Fingerprinting}} & \textbf{Class} \\
		& & V2 & V3.1 & Fingerprint & Cond. for matching\\
		\cmidrule(l{2pt}r{2pt}){3-4} \cmidrule(l{2pt}r{2pt}){5-6}
	}%
	\tablehead{%
		\multicolumn{6}{l}{Continued from previous column} \\
		\toprule
		\textbf{Cl.} & \textbf{CVE} & \multicolumn{2}{r}{\textbf{CVSS}} & \multicolumn{2}{c}{\textbf{Fingerprinting}} & \textbf{Class}\\
		& & V2 & V3.1 & Fingerprint & Cond. for matching\\
		\cmidrule(l{2pt}r{2pt}){3-4} \cmidrule(l{2pt}r{2pt}){5-6}
	}%
	\tabletail{%
	\midrule \multicolumn{6}{l}{{Continued on next column}} \\ \midrule}
	\tablelasttail{%
	\\\midrule
	\multicolumn{6}{l}{{Concluded}} \\ \bottomrule}
	\tablecaption{Overview of vulnerabilities for the \textit{Schneider Electric BMX P34 series}}
	\begin{scriptsize}
		\begin{supertabular*}{\columnwidth}{@{}l l c c c p{0.2\columnwidth} c@{}}
			\label{tab:methodology_vulnerabilities_schneider_electric_overview}
			\centering

			\textbf{1} & \textbf{CVE-2020-7475}  & 9.8 & 7.5 & \Circle & M340 firmware $<$ V3.20 and EcoStruxure Control Expert $<$ V14.1 & D \\
			\midrule
			\multirow{13}{*}{\textbf{2}} & \textbf{CVE-2019-6857}  & 7.5 & 5.0 & \multirow{13}{*}{\CIRCLE} & \multirow{13}{0.2\columnwidth}{M340 firmware $<$ V3.10} & D \\
			& \textbf{CVE-2019-6856} & 7.5 & 5.0 & & & D  \\
			& \textbf{CVE-2018-7794} & 7.5 & 5.0 & & & D \\
			& \textbf{CVE-2019-6829} & 7.5 & 7.8 & & & D \\
			& \textbf{CVE-2019-6828} & 7.5 & 7.8 & & & D \\ 
			& \textbf{CVE-2019-6809} & 7.5 & 7.8 & & & D \\ 
			& \textbf{CVE-2018-7850} & 5.3 & 5.5 & & & T \\
			& \textbf{CVE-2018-7849} & 7.5 & 5.0 & & & D \\ 
			& \textbf{CVE-2018-7848} & 7.5 & 5.0 & & & I \\
			& \textbf{CVE-2018-7847} & 9.8 & 7.5 & & & D/T \\
			& \textbf{CVE-2018-7846} & 9.8 & 5.0 & & & E \\ 
			& \textbf{CVE-2018-7843} & 7.5 & 5.0 & & & D \\
			& \textbf{CVE-2018-7842} & 9.8 & 7.5 & & & E \\
			\midrule
			\multirow{3}{*}{\textbf{3}} & \textbf{CVE-2019-6819} & 7.5 & 5.0 & \multirow{3}{*}{\CIRCLE} & \multirow{3}{0.2\columnwidth}{M340 firmware $<$ V2.90 and specific device} & D \\
			& \textbf{CVE-2018-7851} & 6.5 & 6.8 & & & D \\
			& \textbf{CVE-2018-7845} & 7.5 & 5.0 & & & I \\
			\midrule
			\textbf{4} & \textbf{CVE-2017-6017} & 7.5 & 7.8 & \CIRCLE & M340 firmware $<$ V2.90 and specific device & D \\
			\midrule
			\multirow{9}{*}{\textbf{5}} & \textbf{CVE-2019-6852} & 7.5 & 5.0 & \multirow{9}{*}{\CIRCLE} & \multirow{9}{*}{FTP server running} & I \\
			& \textbf{CVE-2019-6847} & 4.9 & 4.0 & & & D \\
			& \textbf{CVE-2019-6846} & 6.5 & 4.3 & & & I \\
			& \textbf{CVE-2019-6844} & 4.9 & 4.0 & & & D \\
			& \textbf{CVE-2019-6843} & 4.9 & 4.0 & & & D \\ 
			& \textbf{CVE-2019-6842} & 4.9 & 4.0 & & & D \\ 
			& \textbf{CVE-2019-6841} & 4.9 & 4.0 & & & D \\
			& \textbf{CVE-2018-7242} & 7.5 & 5.0 & & & I \\ 
			& \textbf{CVE-2018-7241} & 9.8 & 5.0 & & & I/E \\
			& \textbf{CVE-2018-7847} & 9.8 & 10.0 & & & D/T \\
			\midrule
			\textbf{6} & \textbf{CVE-2019-6851} & 7.5 & 5.0 & \CIRCLE & TFTP Server running & I \\
			\midrule
			\multirow{12}{*}{\textbf{7}} & \textbf{CVE-2019-6845} & 7.5 & 5.0 & \multirow{12}{*}{\CIRCLE} & \multirow{12}{*}{ -- } & I \\
			& \textbf{CVE-2019-6808} & 9.8 & 7.5 & & & T \\
			& \textbf{CVE-2019-6807} & 7.5 & 5.0 & & & D \\
			& \textbf{CVE-2019-6806} & 7.5 & 5.0 & & & I \\
			& \textbf{CVE-2018-7857} & 7.5 & 5.0 & & & D \\ 
			& \textbf{CVE-2018-7856} & 7.5 & 5.0 & & & D \\  
			& \textbf{CVE-2018-7855} & 7.5 & 5.0 & & & D \\ 
			& \textbf{CVE-2018-7854} & 7.5 & 5.0 & & & D \\ 
			& \textbf{CVE-2018-7853} & 7.5 & 5.0 & & & D \\ 
			& \textbf{CVE-2018-7844} & 7.5 & 5.0 & & & I \\ 
			& \textbf{CVE-2018-7852} & 7.5 & 5.0 & & & D \\  
			& \textbf{CVE-2019-6821} & 7.5 & 5.0 & & & I/E \\
			\midrule
			\textbf{8} & \textbf{CVE-2019-6813} & 7.5 & 7.8 & \CIRCLE & SNMP service active & D \\
			\midrule
			\multirow{2}{*}{\textbf{9}} & \textbf{CVE-2018-7833} & 7.5 & 5.0 & \multirow{2}{*}{\CIRCLE} & \multirow{2}{0.2\columnwidth}{M340 firmware $<$ V3.20 and Web server active} & D \\
			& \textbf{CVE-2018-7804} & 6.1 & 5.8 & &   & I \\
			\midrule
			\multirow{10}{*}{\textbf{10}} & \textbf{CVE-2018-7812} & 7.5 & 5.0 & \multirow{10}{*}{\CIRCLE} & \multirow{10}{*}{Web server active} & I \\
			& \textbf{CVE-2018-7831} & 8.8 & 4.3 & & & T/E \\
			& \textbf{CVE-2018-7830} & 7.5 & 5.0 & & & D \\
			& \textbf{CVE-2018-7811} & 9.8 & 5.0 & & & T/E \\
			& \textbf{CVE-2018-7810} & 6.1 & 4.3 & & & T \\ 
			& \textbf{CVE-2018-7809} & 9.8 & 6.4 & & & E \\  
			& \textbf{CVE-2018-7762} & 7.5 & 5.0 & & & D \\ 
			& \textbf{CVE-2018-7761} & 9.8 & 7.5 & & & T/E \\ 
			& \textbf{CVE-2018-7760} & 9.8 & 7.5 & & & E \\ 
			& \textbf{CVE-2018-7759} & 7.5 & 5.0 & & & D \\ 
			\midrule
			\multirow{2}{*}{\textbf{11}} & \textbf{CVE-2015-6462} & 5.4 & 3.5 & \multirow{2}{*}{\CIRCLE} & \multirow{2}{0.2\columnwidth}{Client browser and specific device} & T \\
			& \textbf{CVE-2015-6461} & 5.4 & 5.5 & &   & T \\
			\midrule
			\textbf{12} & \textbf{CVE-2015-7937} & 10.0* & 10.0 & \CIRCLE & GoAhead Web Server and specific device &  T/E \\
			\midrule
			\textbf{13} & \textbf{CVE-2013-2761} & 5.7* & 4.5 & \CIRCLE & FTP server active and specific device & D \\
			\midrule
			\textbf{14} & \textbf{CVE-2014-0754} & 10.0* & 10.0 & \CIRCLE & Web server active and specific device & T/E \\
			\midrule
			\textbf{15} & \textbf{CVE-2011-4859} & 10.0* & 10.0 & \CIRCLE & Telnet or Windriver or FTP active and specific device & E \\
		\end{supertabular*}
	\end{scriptsize}
	\vspace{0.5em}
	
	Since several of the 59 \acp{cve} share the same conditions,
	they are clustered into 15 Clusters in a fashion that they are applicable to the same devices,
	including configuration and firmware version.
	Meaning if a device is susceptible to one \ac{cve} in a cluster,
	it is also susceptible to all the others.
	This is indicated by the leftmost column \textbf{Cl.}.
	It is noteworthy that all vulnerabilities of Cluster 7 solely require access on the Modbus/TCP port in order to be exploitable.
	Since the port was used to discover the devices,
	all devices that are part of the data set are vulnerable. \par
	The table shows that \ac{dos} attacks are the most common,
	followed by a similar number of information disclosure,
	elevation of privilege and tampering attacks.
	The experiments for the \textit{Schneider Electric BMX P34 series} were performed with data queried on April 14th,
	2020.
	A total of 758 was identified matching the criteria presented in Section~\ref{ssec:schneider_electric}.
	The summary of these results is provided in Table~\ref{tab:schneider_overview},
	containing the ten countries hosting the most \textit{Schneider Electric BMX P34 series} devices.
	
	\begin{table}[h!]
		\renewcommand{\arraystretch}{1.3}
		\caption{Overview of vulnerabilities per country for the \textit{Schneider Electric BMX P34 series}}
		\label{tab:schneider_overview}
		\centering
		\scriptsize
		\begin{tabular}{@{}l r r r r r r @{}} 
			\toprule
			\textbf{Country} & \multicolumn{2}{c}{\textbf{No. of Devices}} &  \multicolumn{2}{c}{\textbf{No. of CVEs}} & \multicolumn{2}{c}{\textbf{Weighted by CVSS}}\\
			& Abs.	& Rel. in $\%$ 	& Abs. 	& Rel. in $\%$  & Abs. 	& Rel. in $\%$ \\
			\cmidrule(l{2pt}r{2pt}){2-3} \cmidrule(l{2pt}r{2pt}){4-5} \cmidrule(l{2pt}r{2pt}){6-7} 
			\textbf{France} 		& 255 	& 32.48 	& 1505 	& 32.79 	& 12765.1 & 32.74 \\
			\textbf{Spain} 			& 116 	& 14.78 	& 655 	& 14.27 	&  5230.7 & 13.42 \\
			\textbf{United States} 	& 107 	& 13.63 	& 611 	& 13.31 	&  5531.3 &	14.19 \\
			\textbf{Italy} 			& 74 	& 9.43 		& 410 	& 8.93 		&  3479.5 &  8.94 \\
			\textbf{Turkey} 		& 28 	& 3.57 		& 157 	& 3.42 		&  1337.0 &  3.43 \\
			\textbf{Israel} 		& 22 	& 2.80 		& 136 	& 2.96 		&  1291.1 &  3.31 \\
			\textbf{Canada} 		& 21 	& 2.68 		& 151 	& 3.29 		&  1158.0 &  2.97 \\
			\textbf{Norway} 		& 18 	& 2.29 		& 131 	& 2.85 		&  1102.8 &  2.83 \\
			\textbf{Portugal} 		& 15 	& 1.91 		& 78 	& 1.70 		&   664.0 &  1.70 \\
			\textbf{Poland} 		& 14 	& 1.78 		& 54 	& 1.18 		&   467.1 &  1.19 \\		
			\cmidrule(l{2pt}r{2pt}){2-3} \cmidrule(l{2pt}r{2pt}){4-5} \cmidrule(l{2pt}r{2pt}){6-7} 
			\textbf{$\sum_{Top10}$} & 670 	& 85.35 	& 3888 	& 84.71 	& 33026.6 & 84.72 \\
			\textbf{$\sum_{Total}$} & 785 	& 100 		& 4590 	& 100 		& 38983.9 & 100 \\
			\bottomrule
		\end{tabular}
	\end{table}
	
	In this table,
	$\sum_{Top10}$ is the column sum of the devices used in the ten countries employing most devices,
	while $\sum_{Total}$ is the sum over all devices.
	It can be seen that about 85\% of each sum are found in the top ten countries.
	Most devices were found in France,
	which is the country of origin of Schneider Electric,
	followed by Spain and the United States.
	It can be seen that the relative number of devices per country and the relative number of \acp{cve} per country is within a deviation of .5.
	However,
	the \acp{cve} weighted by the CVSS starts with almost half the relative value,
	compared to the relative value of numbers of \acp{cve} and decreases slower than the number of \acp{cve}.
	That is an indication that,
	although France,
	Spain,
	the United States,
	and Italy operate most devices containing most vulnerabilities,
	these vulnerabilities are not as severe as vulnerabilities of other countries.
	Israel has a slightly higher percentage of the weighted score than of the number of \acp{cve}.
	That means the \acp{cve} the devices operated there are more susceptible to attacks with severe effects. 
	It is noteworthy that,
	since all of the \acp{cve} in Cluster 7,
	as described in Table~\ref{tab:methodology_vulnerabilities_schneider_electric_overview},
	solely rely on Modbus/TCP communication,
	all evaluated devices are vulnerable.
	However,
	there are also vulnerabilities no monitored devices is susceptible to,
	namely CVE-2019-6851 that relies on a running TFTP server and CVE-2014-0754 that requires an active HTTP server as well as a specific device version.
	In general,
	each device is at least susceptible to one CVE,
	at most,
	a device is vulnerable to all clusters except the above-mentioned.
	The mean number of vulnerabilities per device is 5.85.
	Each of the evaluated devices has at least one open port,
	TCP 502,
	which is the default port for Modbus/TCP.
	Apart from that,
	237 devices listen on TCP port 80,
	indicating a web server.
	A running web server is required for exploiting \acp{cve} in Clusters 9,
	10, 11 and 12.
	58 devices listened on TCP port 21,
	commonly used for FTP and required for exploiting \acp{cve} in Clusters 5,
	13,
	and 15.
	54 devices listened on TCP port 23,
	commonly used for the Telnet-protocol.
	It is a remote control protocol that does not provide encryption and is thus only suited for private networks,
	if at all.
	Telnet is required for exploiting \acp{cve} in Clusters 15,
	only containing CVE-2011-4859 which is nine years old.
	In total,
	76 devices are vulnerable to CVE-2011-4859,
	which is a bad sign since the disclosed vulnerability is available for a sufficiently long time. 
	\textit{Shodans} built-in \ac{cve} matcher found a total of 1412 CVEs.
	However,
	the distribution is skewed,
	with a mean of 1.8 \acp{cve} per device,
	a minimum of 0,
	a maximum of 169 and a variance of 124.41.
	That means few devices contain most of the vulnerabilities \textit{Shodan} matched automatically.
	This is unexpected behaviour,
	indicating either devices with several old services,
	invalid \ac{cve} detection of \textit{Shodan},
	or honeypots.  \par
	Additionally,
	the overall distribution of the \acp{cve} found is listed in Table~\ref{tab:schneider_overview_cves}.	
	\begin{table}[h!]
		\renewcommand{\arraystretch}{1.3}
		\caption{Overview of vulnerabilities in the \textit{Schneider Electric BMX P34 series}}
		\label{tab:schneider_overview_cves}
		\centering
		\scriptsize
		\begin{tabular}{l r r r c}
			\toprule
			\textbf{\ac{cve} or Cluster} & \phantom{a} & \multicolumn{2}{r}{\textbf{No. of occurrences}} & \textbf{Class} \\
			&             &       Abs.      & Rel. in $\%$  \\
			\cmidrule{3-4} 
			\textbf{Cluster 7} & & 785 & 100.00 & T/I/D/E \\
			\textbf{CVE-2017-6017} & & 757 & 96.43 & D \\	
			\textbf{Cluster 2} & & 732 & 93.25 & T/I/D/E \\
			\textbf{Cluster 3} & & 712 & 90.70 & I/D \\
			\textbf{Cluster 11} & & 344 & 43.82 & T \\
			\textbf{CVE-2020-7475} & & 314 & 40.00 & D \\
			\textbf{Cluster 5} & & 84 & 10.70 & T/I/D/E \\			
			\textbf{CVE-2011-4859} & & 76 & 9.68 & E \\			
			\textbf{CVE-2013-2761} & & 58 & 7.39 & D \\			
			\textbf{CVE-2019-6813} & & 54 & 6.88 & D \\			
			\textbf{CVE-2019-6851} & & 0 & 0.00 & I \\			
			\textbf{CVE-2014-0754} & & 0 & 0.00 & T/E \\
			\bottomrule			
		\end{tabular}
	\end{table}
	Each device of the \textit{Schneider Electric BMX P34 series} is susceptible to the \acp{cve} in Cluster 7.
	The \acp{cve} in Cluster 7 have a \ac{cvss} 3.1 score of 7.5,
	except for CVE-2019-6808,
	which as a \ac{cvss} 3.1 score of 9.8 and is thus critical.
	This \ac{cve} allows \ac{rce} via Modbus.
	The other \acp{cve} have \ac{cvss} scores that are considered to be high.
	All of them are exploitable via the Modbus protocol,
	consequently finding them in the public network is a critical security issue.
	The remaining \acp{cve} in Cluster 7 either allow a \ac{dos} or the leakage of sensitive information from the device.
	CVE-2017-6017 has a \ac{cvss} 3.1 score of 7.5,
	which is considered high.
	It is only applicable to a certain list of devices and to certain firmware versions.
	However,
	a total of 757 devices,
	or 96.43\% of analysed devices,
	are vulnerable.
	This \ac{cve} allows a \ac{dos} condition that has to be reset by manually pressing a button on the device by an operator.
	732 devices are susceptible to the \acp{cve} in Cluster 2.
	This cluster contains three \acp{cve},
	CVE-2018-7842,
	CVE-2018-7846,
	and CVE-2018-7847,
	with a \ac{cvss} 3.1 score of 9.8,
	which is considered critical.
	These \acp{cve} allow code exection resulting in unauthorised access and elevation of privileges,
	which enables an attacker to directly impact the system.
	CVE-2018-7850 has a \ac{cvss} 3.1 score of 3.5,
	which is considered medium.
	This \ac{cve} allows for the displaying of incorrect information in the associated Unity Pro software.
	All other \acp{cve} in this cluster have a \ac{cvss} 3.1 score of 7.5,
	which is considered high.
	These \acp{cve},
	except for CVE-2018-7848 Detail,
	create \ac{dos} conditions.
	CVE-2018-7848 Detail enables the extraction of information.
	The types of attacks that are evaluated in this scenario describe \ac{rce} attacks,
	data leakage and \ac{dos} attacks.
	\ac{rce} on industrial devices allows an attacker to influence the \ac{cps} connected to the \ac{plc}.
	That means an attacker potentially has the option to misuse \acp{cps} that influence the physical world.
	Causing physical damage to materials and products as well as injuring operators is consequently possible.
	Information leakage can provide business secrets to an attacker,
	which could be stolen due to monetary gain.
	\ac{dos} can be used to disrupt often highly dependant,
	sequential production environments,
	causing a halt in production which often leads to drastic monetary loss.

	\subsection{\textit{Siemens S7-300 series}}
	The Siemens Simatic S7-300 is part of the Siemens Simatic product series.
	This series is the most popular product for process control,
	of which the Simatic S7-300 is the low-end device with limited computational capabilities.
	It was fingerprinted by searching the \textit{Shodan} database for devices satisfying the following two conditions:
	First, the S7 communication port (TCP/102) was exposed to the internet at \textit{Shodan} scan time.
	Second, the received data from that scan contained the string \textit{PLC name: SIMATIC 300},
	which indicates that a correct header of the Simatic S7 product family was received as well as that the field containing the product family is indicating a S7-300 device.
	The resulting \textit{Shodan} query was \textit{port:102 "PLC name: SIMATIC 300"}.
	There are known honeypots supporting the S7 protocol, e.g. Conpot.
	In parallel to the fingerprinting process,
	16 relevant \acp{cve} have been identified for the Simatic S7-300 series.
	The set of \ac{cve} was obtained by searching the \ac{nvd} provided by the \ac{nist} for the key word \textit{SIMATIC S7-300}.
	The set of \acp{cve} was then used to compile a set of fingerprints to identify the subset of vulnerabilities that are applicable to a particular system.
	An overview of the result is given in Table~\ref{tab:methodology_vulnerabilities_siemens_overview}.
	
	\begin{table}[h!]
		\renewcommand{\arraystretch}{1.3}
		\caption{Overview of vulnerabilities of the \textit{Siemens Simatic S7-300 products family}. \CIRCLE: yes, \LEFTcircle: partially, \Circle: no}
		\label{tab:methodology_vulnerabilities_siemens_overview}
		\centering
		\scriptsize
		\begin{tabular}{l r r c p{0.125\textwidth} c}
			\toprule
			\textbf{CVE} & \multicolumn{2}{c}{\textbf{CVSS}} &  \multicolumn{2}{c}{\textbf{Fingerprinting}} & \textbf{Class}\\
			 & V2 & V3.1 & Fingerprint & Cond. for matching & \\
			\cmidrule(l{2pt}r{2pt}){2-3} \cmidrule(l{2pt}r{2pt}){4-5}
			\textbf{CVE-2019-19300} & 5.0 & 7.5 & \CIRCLE & \textit{CPU} in module type & D \\
			\textbf{CVE-2019-18336} & 7.8 & 7.5 & \CIRCLE & \textit{CPU} in module type + firmware $<$ 3.X.17 & D \\
			\textbf{CVE-2019-13940} & 5.0 & 7.5 & \CIRCLE & \textit{PN} or \textit{DP} in module type & D \\
			\textbf{CVE-2019-10936} & 5.0 & 7.5 & \CIRCLE & \textit{CPU} in module type & D \\
			\textbf{CVE-2019-10923} & 5.0 & 7.5 & \CIRCLE & \textit{CPU} in module type & D \\
			\textbf{CVE-2019-6568} & 5.0 & 7.5 & \CIRCLE & \textit{CPU} in module type & D \\
			\textbf{CVE-2018-16561} & 7.8 & 7.5 & \CIRCLE & \textit{CPU} in module type + firmware $<$ 3.X.16 & D \\
			\textbf{CVE-2018-4843} & 6.1 & 6.5 & \CIRCLE & Firmware $<$ 3.X.16 & D \\
			\textbf{CVE-2017-12741} & 7.8 & 7.5 & \CIRCLE & Firmware $<$ 3.X.16 & D \\
			\textbf{CVE-2017-2681} & 6.1 & 6.5 & \CIRCLE & Firmware $<$ 3.X.14 & D \\
			\textbf{CVE-2017-2680} & 6.1 & 6.5 & \CIRCLE & Firmware $<$ 3.X.14 & D \\
			\textbf{CVE-2016-9159} & 4.3 & 5.9 & \CIRCLE & \textit{CPU} in module type & I \\
			\textbf{CVE-2016-9158} & 7.8 & 7.5 & \CIRCLE & \textit{CPU} in module type & D \\
			\textbf{CVE-2016-8673} & 6.8 & 8.8 & \CIRCLE & \textit{PN} or \textit{DP} in module type & E \\
			\textbf{CVE-2016-8672} & 5 & 5.3 & \CIRCLE & \textit{PN} or \textit{DP} in module type & I \\
			\textbf{CVE-2016-3949} & 7.8 & 7.5 & \LEFTcircle & \textit{CPU} in module type $+$ firmware $<$ 3.2.12 or firmware $<$ 3.3.12 & D\\
			\textbf{CVE-2015-2177} & 7.8 & 7.5* & \CIRCLE & \textit{CPU} in module type & D\\
			\bottomrule
		\end{tabular}
	\end{table}

	As it can be seen,
	most \acp{cve} are fingerprintable from the data received from the \textit{Shodan} database.
	\textit{CVE-2016-8672} is an exception,
	as it is not possible to determine if a Simatic S7-300 is Profinet-enabled or Profinet-disabled,
	as Profinet communications and banners might not be exposed to the Internet. \par
	The table shows that most \acp{cve} result in \ac{dos} conditions,
	two allow information disclosure, 
	one allows the elevation of privileges.
	A total of 439 devices were fingerprinted by the signature presented previously.
	Each device has a unique IP address.
	Comparing the geo-spatial distribution of Simatic S7-300 devices found,
	Germany and Italy amount to approximately 16$\%$ each.
	The third ranked country is Spain with around 7$\%$ of the identified devices.
	Interestingly,
	the vulnerabilities are distributed rather different.
	There are only three frequencies with which vulnerabilities occur:
	Six out of 17 occur at \textgreater 99 $\%$,
	three at 80 $\%$ probability and seven occur not at all.
	Out of 439 devices,
	436 were identified to have at least one Siemens Simatic S7-300-specific vulnerability.
	In total 3660 vulnerabilities were found,
	the median of vulnerabilities per device was determined to be 9.
	In Table~\ref{tab:evaluation_vulnerabilities_siemens_overview},
	a comparative overview of the results is presented.
	
	\begin{table}[h!]
		\renewcommand{\arraystretch}{1.3}
		\caption{Overview of vulnerabilities per country for the \textit{Siemens Simatic S7-300 series}}
		\label{tab:evaluation_vulnerabilities_siemens_overview}
		\centering
		\scriptsize
		\begin{tabular}{@{}p{.17\linewidth} r r r r r r @{}} 
			\toprule
			\textbf{Country} & \multicolumn{2}{c}{\textbf{No. of Devices}} &  \multicolumn{2}{c}{\textbf{No. of CVEs}} & \multicolumn{2}{c}{\textbf{Weighted by CVSS}}\\
			& Abs.	& Rel. in $\%$ 	& Abs. 	& Rel. in $\%$  & Abs. 	& Rel. in $\%$ \\
			\cmidrule(l{2pt}r{2pt}){2-3} \cmidrule(l{2pt}r{2pt}){4-5} \cmidrule(l{2pt}r{2pt}){6-7} 
			\textbf{Germany}			& 73 & 16.63	& 690 & 16.85 	& 5004.0 & 16.88 \\
			\textbf{Italy} 				& 71 & 16.17	& 685 & 16.72 	& 4967.0 & 16.76 \\
			\textbf{Spain} 				& 30 &  6.83	& 285 &  6.96 	& 2067.0 &  6.97 \\ 	
			\textbf{Russian Federation}	& 23 &  5.24 	& 230 &  5.62 	& 1667.5 &  5.63 \\
			\textbf{France}				& 22 &  5.01 	& 178 &  4.35 	& 1292.6 &  4.36 \\
			\textbf{Czech Republic}		& 17 &  3.87 	& 164 &  4.00 	& 1189.3 &  4.01 \\
			\textbf{Israel}				& 16 &  3.64 	& 157 &  3.83 	& 1138.4 &  3.84 \\
			\textbf{Poland}				& 16 &  3.64 	& 139 &  3.39 	& 1008.8 &  3.40 \\
			\textbf{Netherlands}		& 14 &  3.19	& 101 &  2.47 	&  734.2 &  2.48 \\		
			\textbf{China}				& 11 &  2.51 	& 110 &  2.69 	&  797.5 &  2.69 \\
			\cmidrule(l{2pt}r{2pt}){2-3} \cmidrule(l{2pt}r{2pt}){4-5} \cmidrule(l{2pt}r{2pt}){6-7} 
			\textbf{$\sum_{Top10}$}		& 293 & 66.74	& 2739 & 66.87 	& 19866.3 & 67.03 \\
			\textbf{$\sum_{Total}$}		& 439 & 100.00	& 4096 & 100.00 & 29636.7 & 100.00 \\
			\bottomrule
		\end{tabular}
	\end{table}

	It can be seen that there is no significant difference between the values with and without the CVSS weighting.
	This is expected behaviour as the corresponding CVSS values presented in Table \ref{tab:methodology_vulnerabilities_siemens_overview} are often exactly $7.5$ or similar.
	A second observation is that the top 10 accounts for roughly 70 $\%$ of the total vulnerabilities found,
	thus following a power distribution.
	This is a common distribution in information security-related statistics \cite{Fraunholz.2017e,Fraunholz.2017h}.
	As there are more factors,
	e.g. insecure configuration,
	that can lead to system vulnerabilities,
	two more parameters are considered to gather a more complete picture of the overall state of security.
	First,
	the number of ports exposed to the internet is enumerated for each identified device.
	Port 102 was found open on each device,
	as it was part of the query.
	The subsequent ports in terms of frequency are 80 (26$\%$) and 443 (22$\%$),
	both are standardised for HTTP and HTTPS web services respectively by IANA.
	Web-based applications are often used to administrate a system or monitor the current state.
	An exposition may result in a compromise or information leakage.
	Followed by the ports 5900 (13$\%$),
	161 (9$\%$)  and 5800 (7$\%$).
	Ports 5800 and 5900 are standardised by IANA for the VNC protocol,
	which provides remote access to a device.
	This remote access may be susceptible to credential guessing-attacks,
	which can result a full system compromise.
	Another potential vulnerability is the exposition of port 21 (6$\%$),
	which is typically used for FTP.
	FTP can be used to upload files,
	e.g. web shells or other backdoors,
	or download potentially sensitive information if the configuration allows this type of access.
	FTP,
	as well as HTTP,
	are inherently susceptible to man-in-the-middle attacks,
	as they to not provide encryption or integrity.
	To further get a picture of the vulnerabilities,
	\textit{Shodan's} vulnerability fingerprinting is used.
	The first interesting observation is that there is no overlap between the \acp{cve} identified by \textit{Shodan} and the \acp{cve} identified by the signatures presented in Table \ref{tab:methodology_vulnerabilities_siemens_overview}.
	Therefore,
	both vulnerabilities sets can be combined to assess the overall security.
	Comparing the absolute numbers,
	reveals that \textit{Shodan} is able to identify 34 devices that are susceptible to at least one vulnerability.
	In total, 779 \acp{cve} of which 305 are unique are identified.
	Curiously,
	there is one single device which is identified to expose 152 vulnerabilities by \textit{Shodan}.
	On average,
	there are 1.8 vulnerabilities per device. \par
	Additionally,
	the overall distribution of the \acp{cve} found is listed in Table~\ref{tab:siemens_overview_cves}.
	
	\begin{table}[h!]
		\renewcommand{\arraystretch}{1.3}
		\caption{Overview of vulnerabilities in the \textit{Siemens S7-300 series}}
		\label{tab:siemens_overview_cves}
		\centering
		\scriptsize
		\begin{tabular}{l r r r c}
			\toprule
			\textbf{\ac{cve}} & \phantom{a} & \multicolumn{2}{c}{\textbf{No. of occurrences}} & \textbf{Class}\\
			&             &       Abs.      & Rel. in $\%$  \\
			\cmidrule{3-4} 
			\textbf{CVE-2019-19300} & & 444 & 100.00 & D \\
			\textbf{CVE-2019-10936} & & 444 & 100.00 & D \\	
			\textbf{CVE-2019-10923} & & 444 & 100.00 & D \\
			\textbf{CVE-2019-6568} & & 444 & 100.00 & D \\
			\textbf{CVE-2016-9159} & & 444 & 100.00 & I \\
			\textbf{CVE-2016-9158} & & 444 & 100.00 & D \\
			\textbf{CVE-2015-2177} & & 444 & 100.00 & D \\
			\textbf{CVE-2019-13940} & & 351 & 79.05 & D \\			
			\textbf{CVE-2016-8673} & & 351 & 79.05 & E \\			
			\textbf{CVE-2016-8672} & & 351 & 79.05 & I \\	
			\bottomrule			
		\end{tabular}
	\end{table}

	The table shows that seven \acp{cve} can be exploited on all vulnerable devices that were found.
	Furthermore, 
	seven \acp{cve} could not be exploited on any device and are thus not listed.
	From the seven \acp{cve},
	all except for CVE-2016-9159 create \ac{dos} situations.
	CVE-2016-9159 in contrast allows an attacker to extract credentials from the device.
	It is assigned a \ac{cvss} 3.1 score of 5.9,
	which is considered medium.
	Every other \ac{cve} is assigned a \ac{cvss} 3.1 score of 7.5,
	which is considered high;
	except for CVE-2015-2177,
	which is not assigned a \ac{cvss} 3.1 score due to its age,
	but has the same \ac{cvss} 2.0 score as the other \acp{cve}.
	The \ac{dos} attacks can be exploited by an attacker with access to the network the \acp{plc} are located in.
	As they can be found by an Internet-based search engine,
	\textit{Shodan},
	they are obviously reachable from the Internet.
	Similarly,
	the credentials can be extracted if an attacker has access to the network.
	In general,
	the most frequent vulnerabilities have a strong impact on the network environment.
	Disrupting process environments can cause a significant loss of money,
	can render materials useless and thus harm an organisation.
	The fact that \acp{cve} that can be exploited just with network access can be discovered is concerning,
	especially since the oldest is from 2015.
	As discussed,
	honeypots could distort the results,
	however,
	it is sufficiently unlikely that all results are honeypots,
	due to their heterogeneity in spatial distribution,
	services running and other characteristics.

	\subsection{\textit{Omron CJ and CS PLC series}}
	The Omron CJ and CS PLC series are part of the Omron PLC product family for process control.
	Other Omron PLC product groups are the CV-series,
	the C200-series,
	the CVM1 and the CQM1H.
	The Omron CJ series was chosen as initial scan results indicated a high prevalence (\textgreater 30$\%$) compared to the other Omron products in the \textit{Shodan} database.
	As the identified vulnerabilities are mostly found in both,
	CJ and CS series,
	the CS series was included as well.
	The series were fingerprinted by searching the \textit{Shodan} database for devices satisfying the following three conditions:
	First, the proprietary Factory Interface Network Service (FINS) protocol port (TCP/9600) was exposed to the internet at \textit{Shodan} scan time.
	Second, the received data from that scan contained the string \textit{response code}.
	As \textit{Shodan} does not support partial matching of strings separated by space,
	the \textit{Shodan} query used to identify Omron PLC devices was \textit{port:9600 "response code"}.
	The third condition was then locally applied to filter for the CJ and CS series by matching the controller model field against the strings \textit{CJ} and \textit{CS}.
	By this process the set of relevant devices was extracted from the \textit{Shodan} database.
	In addition to the fingerprinting process,
	30 \acp{cve} for Omron products have been identified.
	As with the Simatic devices,
	the set of \acp{cve} was obtained by searching the \ac{nvd} of the \ac{nist} for the key word \textit{Omron}.
	Most \acp{cve} are for the Omron CX product line,
	which includes the devices used to program Omron PLCs.
	Even though a compromised programming device is a suitable attack vector,
	only \acp{cve} directly affecting the PLC devices are considered in this study.
	The final set consists of seven \ac{cve} from between 2015 and 2020.
	The set of \acp{cve} was then used to compile a set of fingerprints to identify the subset of vulnerabilities that are applicable to a particular system.
	An overview of the result is given in Table \ref{tab:methodology_vulnerabilities_omron_overview}.
	
	\begin{table}[h!]
		\renewcommand{\arraystretch}{1.3}
		\caption{Overview of vulnerabilities of the \textit{Omron CJ and Cs series}. \CIRCLE: yes, \LEFTcircle: partially, \Circle: no}
		\label{tab:methodology_vulnerabilities_omron_overview}
		\centering
		\scriptsize
		\begin{tabular}{@{}l r r c p{0.14\textwidth} c@{}}
			\toprule
			\textbf{CVE} & \multicolumn{2}{r}{\textbf{CVSS}} & \multicolumn{2}{c}{\textbf{Fingerprinting}} & \textbf{Class}\\
			&                      V2 & V3.1                                &        Fingerprint & Cond. for matching &\\
			\cmidrule(l{2pt}r{2pt}){2-3} \cmidrule(l{2pt}r{2pt}){4-5}
			\textbf{CVE-2020-6986} & 7.8 & 7.5 & \CIRCLE & \textit{CJ} in controller model & D\\
			\textbf{CVE-2019-18269} & 7.5 & 9.8 & \CIRCLE & \textit{CJ} in controller model & T/D/E \\
			\textbf{CVE-2019-18261} & 5.0 & 9.8 & \CIRCLE & \textit{CJ} or \textit{NJ} or \textit{CS} in controller model & T/E \\
			\textbf{CVE-2019-18259} & 7.5 & 9.8 & \CIRCLE & \textit{CJ} or \textit{CS} in controller model & T/I \\
			\textbf{CVE-2019-13533} & 6.8 & 8.1 & \CIRCLE & \textit{CJ} or \textit{CS} in controller model & T \\
			\textbf{CVE-2015-1015} & 2.1 & 4.0* & \CIRCLE & \textit{CJ2M} in con. mod. $+$ con. ver. \textless 2.1 or \textit{CJ2H} in con. mod. $+$ con. ver. \textless 1.5 & I \\
			\textbf{CVE-2015-0987} & 5.0 & 5.3* & \CIRCLE & \textit{CJ2M} in con. mod. $+$ con. ver. \textless 2.1 or \textit{CJ2H} in con. mod. $+$ con. ver. \textless 1.5 & I \\
			\bottomrule
		\end{tabular}
	\end{table}

	As it can be seen,
	each \ac{cve} is fingerprintable from the data received from the \textit{Shodan} database.
	This enables a broad insight into the vulnerabilities that are prevalent in the Omron CJ and CS series PLCs exposed to the Internet. \par
	A total of 1579 devices was fingerprinted according to the previously presented signature.
	Each device has a unique IP address.
	Comparing the geo-spatial distribution of Omron CJ and CS PLC found,
	Spain is the origin of approximately 25$\%$ of them.
	The second and third ranked countries are France and Canada each with around 10$\%$ of the identified devices.
	Comparing this distribution the country distribution of the devices with an identified vulnerability,
	it can be observed that both are similar for the top 3 countries.
	However,
	the percentage of devices in Spain is reduced to 19$\%$ when only devices with \acp{cve} are considered.
	This indicates that devices in Spain have a slightly better security compared to the average.
	Five out of seven vulnerabilities occur at about 63$\%$ of all devices identified as Omron product.
	The remaining two vulnerabilities occur at a significantly lower rate of about 15$\%$ of the devices.
	Out of 1579 devices,
	1018 were identified to have at least one Omron-specific vulnerability.
	In total, 5219 vulnerabilities were found,
	the median of vulnerabilities per device was determined to be 5.
	In Table~\ref{tab:evaluation_vulnerabilities_omron_overview},
	a comparative overview of the results is presented.
	

\begin{table}[h!]
	\renewcommand{\arraystretch}{1.3}
	\caption{Overview of vulnerabilities per country for the \textit{Omron CJ and CS PLC series}}
	\label{tab:evaluation_vulnerabilities_omron_overview}
	\centering
	\scriptsize
	\begin{tabular}{l rr rr rr}
		\toprule
		\textbf{Country} & \multicolumn{2}{c}{\textbf{No. of devices}} & \multicolumn{2}{r}{\textbf{No. of CVEs}} &  \multicolumn{2}{r}{\textbf{Weighted by CVSS}}\\
		\cmidrule(l{2pt}r{2pt}){2-3} \cmidrule(l{2pt}r{2pt}){4-5} \cmidrule(l{2pt}r{2pt}){6-7}
		\textbf{Spain}          & 392 & 24.83 & 978 & 18.74 & 8702.7 & 19.02 \\
		\textbf{France}         & 165 & 10.45 & 547 & 10.48 & 4763.0 & 10.41 \\
		\textbf{Canada}         & 152 & 9.63 & 557 & 10.67 & 4951.2 & 10.82 \\
		\textbf{United States}  & 119 & 7.54 & 262 & 5.02  & 2327.0 & 5.09 \\
		\textbf{Hungary}        & 89 & 5.64 & 454 & 8.7  & 3756.7 & 8.21 \\
		\textbf{Italy}          & 86 & 5.45 & 432 & 8.28  & 3683.8 & 8.05 \\
		\textbf{Portugal}	    & 79 & 5.0 & 324 & 5.95  & 1988.8 & 4.35 \\
		\textbf{Netherlands}    & 70 & 4.43 & 320 & 6.13  & 2864.9 & 6.26 \\
		\textbf{Belarus}        & 54 & 3.42 & 171 & 3.28  & 1534.0 & 3.35 \\		
		\textbf{Taiwan}         & 49 & 3.1 & 249 & 4.77  &  2146.0 & 4.69 \\
		\cmidrule(l{2pt}r{2pt}){2-3} \cmidrule(l{2pt}r{2pt}){4-5} \cmidrule(l{2pt}r{2pt}){6-7}
		\textbf{$\sum_{Top10}$} & 1255 & 79.48 & 4197 & 80.42 & 36718.1  & 80.25  \\
		\textbf{$\sum_{Total}$} & 1579 & 100 & 5219 & 100 & 45755.3 & 100 \\
		\bottomrule
	\end{tabular}
\end{table}

	It can be seen that there is no significant difference between the values with and without the CVSS weighting.
	This can be an indication that the vulnerabilities are equally distributed among the vulnerable devices and countries.
	A second observation is that the top 10 accounts for more than 80 $\%$ of the total of vulnerabilities found,
	thus following a power distribution.
	This is a common distribution in information security-related statistics \cite{Fraunholz.2017e,Fraunholz.2017h}.
	To complement the overview of the state of security,
	two further factors are considered.
	First,
	the number of ports exposed to the internet is enumerated for each identified device.
	Port 9600 was found open on each device.
	This was expected as one of the two conditions to fingerprint a device as Omron CJ and CS PLC series was the exposition of port 9600.
	The subsequent ports in terms of frequency are 80 (26$\%$) and 443 (18$\%$),
	both are standardised for web services, namely HTTP and HTTPS, by IANA.
	Web-based applications are often used to administrate a system or monitor the current state.
	An exposition may result in a full compromise or information leakage scenario.
	Next in rank is the port 21 (11$\%$),
	which is used for FTP-based file transfer.
	Exposing this port to the internet is a security risk as it has no encryption and integrity protection.
	A compromised FTP service can lead to full system compromise or information disclosure.
	Additionally,
	port 5900 (11$\%$) and 22 (10$\%$) are found open.
	These ports are used for VNC and SSH,
	both of which being remote administration services.
	Having publicly available administration services may be a risk as in many cases no or default credentials are used to secure them.
	Furthermore,
	they are often susceptible to credential-guessing attacks.
	Port 23 (6$\%$) is also a major security risk.
	Port 23 is standardised for Telnet by IANA.
	Telnet,
	as well as FTP,
	does not have any encryption or integrity protection,
	thus being susceptible to man-in-the-middle scenarios.
	A compromised Telnet service grants full access to the system.
	To further analyse the state of security of the Omron CJ and CS PLC series,
	\textit{Shodans} vulnerability fingerprinting is used.
	The first interesting observation is that there is no overlap between the \acp{cve} identified by \textit{Shodan} and the \acp{cve} identified by the signatures presented in Table \ref{tab:methodology_vulnerabilities_omron_overview}.
	Therefore,
	both vulnerabilities sets can be combined to assess the overall security.
	Comparing the absolute numbers,
	reveals that \textit{Shodan} is able to identify 128 devices that are susceptible to at least one vulnerability.
	In total, 2213 \acp{cve} of which 426 are unique are identified.
	Curiously,
	there is one single device which is identified to expose 121 vulnerabilities by \textit{Shodan}.
	In average,
	there are 1.4 vulnerabilities per device. \par
	Additionally,
	the overall distribution of the \acp{cve} found is listed in Table~\ref{tab:omron_overview_cves}.
	\begin{table}[h!]
		\renewcommand{\arraystretch}{1.3}
		\caption{Overview of vulnerabilities in the \textit{Omron CJ and CS PLC series}}
		\label{tab:omron_overview_cves}
		\centering
		\scriptsize
		\begin{tabular}{l r r c}
			\toprule
			\textbf{\ac{cve} or Cluster} &  \multicolumn{2}{c}{\textbf{No. of occurrences}} & \textbf{Class} \\
			&       Abs.      & Rel. in $\%$ &  \\
			\cmidrule{2-3} 
			\textbf{CVE-2019-18261} & 1018 & 64.47 & T/E \\
			\textbf{CVE-2019-18259} & 1001 & 63.39 & T/I \\
			\textbf{CVE-2019-13533} & 1001 & 63.39 & T \\
			\textbf{CVE-2020-6986} & 980 & 62.06 & D \\
			\textbf{CVE-2019-18269} & 980 & 62.06 & T/D/E \\
			\textbf{CVE-2015-1015} & 239 & 15.14 & I \\
			\bottomrule			
		\end{tabular}
	\end{table}

	Several noteworthy features of the \textit{Omron CJ and CS PLC series} can be observed.
	First,
	no \ac{cve} can be attributed to the whole set of devices found.
	Second,
	compared to the \textit{Schneider Electric BMX P34 series} and the \textit{Siemens S7-300 series},
	the \textit{Omron CJ and CS PLC series} is less susceptible to \ac{dos} attacks.
	Instead,
	the \acp{cve} allow for tampering,
	elevation of privilege and information disclosure.
	Especially tampering can have catastrophic consequences in an \ac{ot} environment jand ultimately cause bodily harm.
	CVE-2019-18261 allows for brute force attacks to gain access to the device,
	which extends the attackers circle of influence.
	CVE-2019-18259 allows for an attacker to spoof messages or execute arbitrary commands,
	meaning an attacker with access to the network can actively influence the device and all connected actuators.
	CVE-2019-13533 allows for replay attacks,
	which enable an attacker to tamper with the physical environment,
	again potentially causing catastrophic results.
	CVE-2020-6986 can lead to a \ac{dos} condition,
	disrupting a process,
	and CVE-2019-18269 enables attackers to tamper with locks in the software flow.

	\subsection{\textit{Rockwell Automation/Allen-Bradley MicroLogix 1400 series}}
	The \textit{Rockwell Automation/Allen-Bradley MicroLogix 1400 series} encompasses six different controllers and one memory module. 
	In fingerprinting the series,
	the \textit{Shodan} functionalities were used to make sure that the scanned device is a product of Rockwell Automation and shows a version starting with 1766,
	which is used for the \textit{MicroLogix 1400 series}.
	The main limitation of this fingerprinting method is its reliance on the correct product name and version number being available.
	However,
	circumventing this limitation by instead introducing expected services to the search parameters would yield far too many false positives.
	The resulting search query was \textit{'product:Rockwell Automation' 'version:1766-'}. 
	In order to find relevant \acp{cve},
	the \ac{nvd} of the \ac{nist} was queried for \textit{cpe:2.3:h:rockwellautomation:ab\_micrologix\_controller:1400:} \linebreak \textit{*:*:*:*:*:*:*},
	which uses the most recent edition of the Official Common Platform Enumeration Dictionary \cite{Cichonski.2011}.
	Missing from the query is the firmware version, which was not searchable by \textit{Shodan}. 
	Therefore, a higher rate of false positives is to be expected, since many of the vulnerabilities are mitigated by firmware upgrades. 
	The relevant \acp{cve} can be seen in Table~\ref{tab:methodology_vulnerabilities_rockwell_overview}. 
	
	\begin{table}[h!]
		\renewcommand{\arraystretch}{1.3}
		\caption{Overview of vulnerabilities of the \textit{Rockwell Automation/Allen-Bradley MicroLogix 1400 series}. \CIRCLE: yes, \LEFTcircle: partially, \Circle: no}
		\label{tab:methodology_vulnerabilities_rockwell_overview}
		\centering
		\scriptsize
		\begin{tabular}{l r r c p{0.13\textwidth} c@{}}
			\toprule
			\textbf{CVE} &  \multicolumn{2}{c}{\textbf{CVSS}} & \multicolumn{2}{c}{\textbf{Fingerprinting}} & \textbf{Class}\\
			             &         V2 & V3.1                               &        Fingerprint & Cond. for matching & \\
			\cmidrule(l{2pt}r{2pt}){2-3} \cmidrule(l{2pt}r{2pt}){4-5}
			\textbf{CVE-2017-7903} & 5.0 & 9.8 & \CIRCLE & Series A and B, version 16.00 and prior & S \\
			\textbf{CVE-2017-7902} & 5.0 & 9.8 & \CIRCLE & Series A and B, version 16.00 and prior & T \\
			\textbf{CVE-2017-7901} & 9.0 & 8.6 & \CIRCLE & Series A and B, version 16.00 and prior & S \\
			\textbf{CVE-2017-7899} & 5.0 & 9.8 & \CIRCLE & Series A and B, version 16.00 and prior & S \\
			\textbf{CVE-2017-7898} & 5.0 & 9.8 & \CIRCLE & Series A and B, version 16.00 and prior & S \\
			\textbf{CVE-2014-5410} & 7.1 & 7.5* & \CIRCLE & Series A, version 7.00 and prior; Series B, version 15.001 and prior & D \\
			\textbf{CVE-2012-4690} & 7.1 & 7.5* & \LEFTcircle & MicroLogix controller 1100, 1200, 1400, and 1500 & D\\
			\bottomrule
		\end{tabular}
	\end{table} 

	The queries for the \textit{Rockwell Automation/Allen-Bradley MicroLogix 1400 series} were launched on April 22nd, 2020. 
	A total number of 1832 devices was fingerprinted,
	each with its own unique IP address. 
	With $65\%$, the majority of devices were located in the USA,
	the second and third place being Canada and Portugal with $5-6\%$ each.
	This result is congruent with the calculation of exposed devices by factoring in the vendor-specific \acp{cve} per country. 
	Since not all vulnerabilities are of equal severity,
	Table~\ref{tab:rockwell_overview} additionally shows a weighted score that is the product of the number of \acp{cve} and their \ac{cvss}.
	The weighted ranking is almost identical to the unweighted ranking by number of devices, with the exceptions of Australia and Norway, both of whom have a slightly higher percentage in the weighted ranking. 

\begin{table}[h!]
	\renewcommand{\arraystretch}{1.3}
	\caption{Overview of vulnerabilities per country for the \textit{Rockwell Automation/Allen-Bradley MicroLogix 1400 series}}
	\label{tab:rockwell_overview}
	\centering
	\scriptsize
	\begin{tabular}{@{}l r r r r r r@{}}
		\toprule
		\textbf{Country} & \multicolumn{2}{c}{\textbf{No. of Devices}} &  \multicolumn{2}{c}{\textbf{No. of CVEs}} &  \multicolumn{2}{c}{\textbf{Weighted by CVSS}}\\
		&       Abs.      & Rel. in $\%$                               &        Abs. & Rel. in $\%$  &                     Abs. & Rel. in $\%$ \\
		\cmidrule(l{2pt}r{2pt}){2-3} \cmidrule(l{2pt}r{2pt}){4-5} \cmidrule(l{2pt}r{2pt}){6-7}
		\textbf{USA} & 1197 & 65.37 & 5673 & 62.84 & 46173.3 & 62.88 \\
		\textbf{Canada} & 108 & 5.90 & 619 & 6.86 & 5039.9 & 6.86 \\
		\textbf{Portugal} & 100 & 5.46 &  397 & 4.80 & 3208.7 & 4.37 \\
		\textbf{Australia} & 75 & 4.10 & 405 & 4.4 & 3295.5 & 4.49 \\
		\textbf{Italy} & 61 & 3.33 & 307 & 3.4 & 2487.2 & 3.39 \\
		\textbf{New Zealand} & 55 & 3.00 & 319 & 3.53 & 2594.9 & 3.53 \\
		\textbf{Spain} & 51 & 2.79 & 249 & 2.76 & 2015.4 & 2.74 \\
		\textbf{Norway} & 46 & 2.51 & 310 & 3.43 & 2531.0 & 3.45 \\
		\textbf{Taiwan} & 26 & 1.42 & 124 & 1.37 &  1007.9 & 1.37 \\
		\textbf{United Kingdom} & 22 & 1.20 & 148 & 1.64 & 1208.3 & 1.65 \\		
		\cmidrule(l{2pt}r{2pt}){2-3} \cmidrule(l{2pt}r{2pt}){4-5} \cmidrule(l{2pt}r{2pt}){6-7}
		\textbf{$\sum_{Top10}$} & 1741 & 95.08 & 8551 & 94.72 & 69562.1 & 94.72 \\
		\textbf{$\sum_{Total}$} & 1831 & 100 & 9028 & 100 & 73436.3 & 100 \\
		\bottomrule
	\end{tabular}
\end{table}

	As shown in Table~\ref{tab:methodology_vulnerabilities_rockwell_overview},
	CVE-2012-4690 does not distinguish between versions,
	which is why all of the 1831 scanned devices are flagged as susceptible to the corresponding attack.
	The other six \acp{cve} were found to fit over half of the scanned devices, with percentages ranging from $53.09\%$ to $68.00\%$. 
	In total, 
	9028 vulnerabilities were found. 
	The most \acp{cve} per device found were 7,
	which means that all vendor-specific \acp{cve} were present. 
	On average,
	every device showed 4.4 vulnerabilities. 
	In addition to the vendor-specific \acp{cve} there was also a distinct set of \textit{Shodan} \acp{cve},
	with 81 devices being susceptible to at least one of them. In total,
	2151 of \textit{Shodan} vulnerabilities were found,
	with an average of 1.17 vulnerabilities per device. 
	The scan showed a total number of 5664 open ports and therefore services.
	Naturally,
	not every open port is necessarily a vulnerability,
	though in general, opening up a device to the Internet always increases the attack surface and therefore the risk of damage.	
	
	\begin{table}[h!]
		\renewcommand{\arraystretch}{1.3}
		\caption{Overview of exposed ports and services for the \textit{Rockwell Automation/Allen-Bradley MicroLogix 1400 series}}
		\label{tab:rockwell_overview_ports}
		\centering
		\scriptsize
		\begin{tabular}{l r r r r l}
			\toprule
			\textbf{Port} & \phantom{a} & \multicolumn{2}{c}{\textbf{No. of Ports}} & \phantom{a} & \textbf{Service}\\
			&	& Abs. 	& Rel. in $\%$ 	& 	& 	\\
			\cmidrule{3-4} 		
			44818 	& 	& 1831 	& 100.00 			& 	& EtherNet/IP explicit messaging	\\
			80 		& 	& 502	& 27.42 		& 	& HTTP	\\
			443 	& 	& 447 	& 24.41 		& 	& HTTPS	\\
			9191 	& 	& 249 	& 13.60 			& 	& Sierra Wireless Airlink	\\
			8080 	& 	& 230 	& 12.56 		& 	& HTTP	\\
			1723 	& 	& 210 	& 11.47 		& 	& PPTP 	\\
			9443 	& 	& 161 	& 8.79 			& 	& VMware Websense Triton console	\\
			5900 	& 	& 157	& 8.57 			& 	& RFB	\\
			8443 	& 	& 136 	& 7.43 			& 	& SSl	\\
			10000 	& 	& 122 	& 6.66			&	& NDMP	\\
			\bottomrule
		\end{tabular}
	\end{table}
	
	Table~\ref{tab:rockwell_overview_ports} shows the ten most frequent open ports. 
	By far the most common service found is EtherNet/IP which is both very common for \acp{plc} and a key feature of the \textit{MicroLogix 1400 PLC},
	and was found in every scanned device.
	Over half of the scanned devices showed HTTP or HTTPS ports.
	Interestingly,
	$11.47\%$ of the scanned devices had an open \ac{pptp} port,
	which could mean serious security issues since \ac{pptp} is fundamentally flawed with respect to encryption and authentication.
	\par 
	Table~\ref{tab:rockwell_overview_cves} provides an overview of the prevalence of the vendor-specific \acp{cve} by showing the absolute and relative numbers of devices that were flagged as susceptible to each \ac{cve}. 	
	
	%
	\begin{table}[h!]
		\renewcommand{\arraystretch}{1.3}
		\caption{Overview of vulnerabilities in the \textit{Rockwell Automation/Allen-Bradley MicroLogix 1400 series}}
		\label{tab:rockwell_overview_cves}
		\centering
		\scriptsize
		\begin{tabular}{l r r c}
			\toprule
			\textbf{CVE} & \multicolumn{2}{c}{\textbf{No. of occurrences}} & \textbf{Class} \\
			& Abs. & Rel. in \%  & \\
			\cmidrule{2-3} 
			\textbf{CVE-2012-4690} & 1831 & 100.00 & D \\
			\textbf{CVE-2017-7898} & 1245 & 68.00 & S \\
			\textbf{CVE-2017-7899} & 1245 & 68.00 & S \\
			\textbf{CVE-2017-7901} & 1245 & 68.00 & S \\
			\textbf{CVE-2017-7902} & 1245 & 68.00 & T \\
			\textbf{CVE-2017-7903} & 1245 & 68.00 & S \\
			\textbf{CVE-2014-5410} & 972 & 53.09 & D \\
			\bottomrule
		\end{tabular}
	\end{table}	
	
	CVE-2012-4690 occurs when static status is not enabled, which opens the system up to a \ac{dos} attack, since remote attackers can modify the status bits. The severity of CVE-2012-4690 is high, with a CVSSv3 of 7.5. \par  
	The next five vulnerabilities were all equally common, affecting 68\% of devices. CVE-2017-7898 allows for remote brute force attacks, since the system does not set a limit on repeated authentication attempts. This is made even more precarious by the system's use of numeric passwords of short length, which is described in CVE-2017-7903. Both vulnerabilities are critical. \par 
	CVE-2017-7899 allows a local attacker to obtain authentication credentials by reading the server logs, since the system accepts them being sent by GET requests. With a CVSSv3 of 9.8, this too is a critical vulnerability. \par
	CVE-2017-7901 is of high severity. It allows remote attackers to spoof TCP connections or launch DoS attacks, made possible by insufficiently random initial sequence numbers in the server's TCP communication. \par 
	CVE-2017-7902 opens the system up to replay attacks, and scores as critical. Nonces, which usually mitigate this kind of attack, are reused by the system, which defeats their purpose when faced with an attacker that can monitor the network. \par 
	Lastly, attackers can send malformed packets over Ethernet or \ac{slip} to execute a DoS attack. This vulnerability, described in CVE-2014-5410, has a high severity.

	\subsection{\textit{Mitsubishi Melsec Q series}}
	According to Figure~\ref{fig:vendor_distribution},
	Mitsubishi has a significant share on the PLC market.
	The PLC series of Mitsubishi is called Melsec series.
	A product line of this series,
	the Melsec-Q Series,
	uses a proprietary service protocol operating by default on TCP port 5006 and 5007.
	\textit{Shodan} only lists relatively few of theses devices, between 200 and 300, by the time this paper is written,
	that are exposed to the Internet.
	The devices can be easily identified by the TCP port and banner stating the device model with a string like \textit{CPU: Q03UDECPU} for the respective model.
	Querying the \ac{cve} database of \ac{nist} for the Melsec service protocol results in eight \acp{cve}.
	Of these results,
	only four devices affected by the \acp{cve} can be queried with \textit{Shodan}. 
	As the firmware versions of the respective devices are not listed on \textit{Shodan},
	it is impossible to say whether the devices are patched, so a high false positive rate for the survey of Mitsubishi devices should be expected.
	The \acp{cve} that have been considered are from the years 2019 and 2020.
	The vulnerabilities CVE-2019-10977 and CVE-2016-8370 affect the Melsec-Q Ethernet interface module which could not be detected by \textit{Shodan} and was therefore not taken into account for this work.
	The CVE-2019-13555 affects the FTP server of the PLC by default running on port 21 and returning the banner \textit{"220 iQ-R FTP server ready."}.
	Searching \textit{Shodan} for the banner reveals 44 devices, but it is likely that these are honeypots, which will be discussed in detail in Section~\ref{sec:discussion}.	
	
	\begin{table}[h!]
		\renewcommand{\arraystretch}{1.3}
		\caption{Overview of vulnerabilities of the \textit{Mitsubishi Melsec Q Series}. \CIRCLE: yes, \LEFTcircle: partially, \Circle: no}
		\label{tab:methodology_vulnerabilities_mitsubishi_overview}
		\centering
		\scriptsize
		\begin{tabular}{@{}l r r c p{0.125\textwidth} c@{}}
			\toprule
			\textbf{CVE} & \multicolumn{2}{c}{\textbf{CVSS}} & \multicolumn{2}{c}{\textbf{Fingerprinting}} & \textbf{Class}\\
			&                     V2 & V3.1                                &        Fingerprint & Cond. for matching & \\
			\cmidrule(l{2pt}r{2pt}){2-3} \cmidrule(l{2pt}r{2pt}){4-5}
			\textbf{CVE-2020-5527} &  5.0 & 7.5  & \LEFTCIRCLE & Ports and Productname & D\\
			\textbf{CVE-2019-13555}  & 4.3 & 5.9  & \LEFTCIRCLE & CPU name in banner and port 21 (FTP) & D\\
			\textbf{CVE-2019-6535}  & 5.0 & 7.5  & \LEFTcircle & CPU name in banner & D\\
			\bottomrule
		\end{tabular}
	\end{table}
	
	The \textit{Shodan} query for Mitsubishi Melsec devices produced 259 results,
	of which 187 were unique IP addresses. The duplicates are listed once for the open UDP port 5006 and once for the open TCP port 5007.
	In this evaluation only unique IP addresses were considered as one device. 
	The Melsec Q devices were found most often in Japan, the country of origin of Mitsubishi, with 119 devices or 63.3$\%$.
	Second is USA with 15 devices and Poland with 8.
	Of the CVEs considered in Table~\ref{tab:methodology_vulnerabilities_mitsubishi_overview},
	173 vulnerabilities were found in 119 devices of Japan and 30 in the devices of the USA.
	It is not clear whether the vulnerabilities are actually present for the firmware of the devices, as only the model was fingerprinted.
	For this evaluation it is assumed no device is patched,
	which is the most conservative assumption.
	The distribution of devices and of \acp{cve} over the countries is nearly even. 
	The mean of CVEs per device was found to be 1.56 with a median of CVEs per device of 2.0.
	In Table~\ref{tab:mitsubishi_overview}, an overview of the results is provided.	
	
	\begin{table}[h!]
		\renewcommand{\arraystretch}{1.3}
		\caption{Overview of vulnerabilities per country for \textit{Mitsubishi Melsec devices}}
		\label{tab:mitsubishi_overview}
		\centering
		\scriptsize
		\begin{tabular}{@{}l r r r r r r @{}} 
			\toprule
			\textbf{Country} & \multicolumn{2}{c}{\textbf{No. of Devices}} &  \multicolumn{2}{c}{\textbf{No. of CVEs}} & \multicolumn{2}{c}{\textbf{Weighted by CVSS}}\\
			& Abs.	& Rel. in $\%$ 	& Abs. 	& Rel. in $\%$  & Abs. 	& Rel. in $\%$ \\
			\cmidrule(l{2pt}r{2pt}){2-3} \cmidrule(l{2pt}r{2pt}){4-5} \cmidrule(l{2pt}r{2pt}){6-7} 
			\textbf{Japan} 		& 119 	& 63.30 & 173 	& 59.04 & 1289.5& 58.98 \\
			\textbf{USA} 		& 15 	& 7.98 	& 30 	& 10.24 & 225.0	& 10.29 \\
			\textbf{Poland} 	& 8 	& 4.26 	& 15 	& 5.12 	& 112.5	& 5.15 \\
			\textbf{South Korea}& 6 	& 3.19 	& 12 	& 4.10 	& 90.0 	& 4.12 \\
			\textbf{Thailand} 	& 4 	& 2.13 	& 8 	& 2.73 	& 60.0 	& 2.74 \\
			\textbf{Canada} 	& 5 	& 2.66 	&  7 	& 2.39 	& 52.5 	& 2.40 \\
			\textbf{Sweden} 	& 5 	& 2.66 	& 7 	& 2.39 	& 52.5 	& 2.40 \\
			\textbf{Norway} 	& 3 	& 1.60 	& 6 	& 2.05 	&  45.0 & 2.06 \\
			\textbf{Spain} 		& 2 	& 1.06 	& 6 	& 2.05 	& 41.8 	& 1.91 \\
			\textbf{United Kingdom} & 3 & 1.60 	& 5 	& 1.71 	& 37.5 	& 1.72 \\	
			\cmidrule(l{2pt}r{2pt}){2-3} \cmidrule(l{2pt}r{2pt}){4-5} \cmidrule(l{2pt}r{2pt}){6-7} 
			\textbf{$\sum_{Top10}$} &170 & 90.44& 269 	& 91.81 & 2006.3 & 91.77  \\
			\textbf{$\sum_{Total}$} & 188 & 100 & 293 	& 100 	& 2186.3 & 100 \\
			\bottomrule
		\end{tabular}
	\end{table}
	
	Weighting the CVEs with the CVSS 3 score does not change the relative distribution of impact per country as the CVSS 3 score for the most common vulnerabilities are equal. 
	Since only in the top five countries more than five devices were found, the top ten countries account for 90,44$\%$ of all devices.
	Either TCP port 5007 or UDP port 5006 have to be open on each device, as these were part of the search query.
	These ports are used for the custom networking protocol on the Melsec devices.
	86 devices has port 80 open for which is assigned the HTTP protocol by the IANA.
	Also frequently open were port 23 (32.45$\%$) for Telnet, a remote terminal, and port 21 (28.72$\%$) for FTP, a protocol used for file transfer.
	Both Telnet and FTP pose a security risk as they do not encrypt communication and no integrity checks and therefore allow man-in-the-middle attacks.
	Less frequently the ports 8080, 9191, 590, 2332 and 5009 are found to be open.
	The \textit{Shodan} vulnerability analysis found 10 CVEs of which none overlap with the CVEs identified in Table \ref{tab:methodology_vulnerabilities_mitsubishi_overview}.
	\textit{Shodan} only found CVEs matching 9 devices which are mostly located in Poland, the UK and France.
	Combining the \textit{Shodan} results with the vendor specific vulnerabilities increases the vulnerabilities found on devices in Poland from 15 to 62.
	In total, \textit{Shodan} identified 114 CVEs, of which 57 are unique.
	Note that many of the Melsec Q devices located in Japan have an IP address assigned to "Research Organization of Information and Systems" and may be exposed to the internet intentionally to study potential attacks,
	as described in Section~\ref{sec:discussion}.
				
    Additionally, the overall distribution of the CVEs found is listed in Table~\ref{tab:mitsubishi_overview_cves}.
	 \begin{table}[h!]
		\renewcommand{\arraystretch}{1.3}
		\caption{Overview of vulnerabilities in the \textit{Mitsubishi Melsec Q Series}}
		\label{tab:mitsubishi_overview_cves}
		\centering
		\scriptsize
		\begin{tabular}{l r r c}
			\toprule
			\textbf{CVE} & \multicolumn{2}{c}{\textbf{No. of occurrences}} & \textbf{Class} \\
			& Abs. & Rel. in \% & \\
			\cmidrule{2-3} 
			\textbf{CVE-2020-5527} & 188 & 100.00 & D \\
			\textbf{CVE-2019-6535} & 98 & 52.13 & D \\
			\textbf{CVE-2019-13555} & 7 & 3.72 & D\\
			\bottomrule
		\end{tabular}
	\end{table}
    All three CVEs create DoS conditions which can be used to disrupt often highly dependant, sequential production environments, causing a halt in production, as previously described.
	CVE-2020-5527 was assumed to apply to all 188 devices, CVE-2019-6535 applies to 52.13$\%$ of the devices with 98 occurrences and CVE-2019-13555 to 7 devices.
	CVE-2019-13555 requires the FTP port 21 to be open and certain CPU-types,
	resulting in the low distribution.	
	Table~\ref{tab:device-comparison} shows a comparison of all devices on a set of important metrics.
		
	\begin{table*}
		\caption{Inter-device comparison}
		\label{tab:device-comparison}
		\centering
		\scriptsize
		\begin{tabular}{@{}p{0.16\linewidth}@{\hspace{0.5em}} rr rrrrrr rr rr r}
			\toprule
			\multicolumn{1}{c}{\textbf{Device type}} & \multicolumn{2}{c}{\textbf{No. of Devices}} & \multicolumn{6}{c}{\textbf{Class} (rel. in \%)} & \multicolumn{2}{c}{\textbf{No. of CVEs}} & \multicolumn{2}{c}{\textbf{Weighted by CVSS}} & \textbf{$\bar{x}$ CVSS}\\
			 & \multicolumn{1}{c}{Abs.} & \multicolumn{1}{c}{Rel. (\%)} & \multicolumn{1}{c}{S} & \multicolumn{1}{c}{T} & \multicolumn{1}{c}{R} & \multicolumn{1}{c}{I} & \multicolumn{1}{c}{D} & \multicolumn{1}{c}{E} & \multicolumn{1}{c}{Abs.} & \multicolumn{1}{c}{Rel. (\%)} & \multicolumn{1}{c}{Abs.} & \multicolumn{1}{c}{Rel. (\%)} & \\
			 \cmidrule(l{2pt}r{2pt}){2-3} \cmidrule(l{2pt}r{2pt}){4-9} \cmidrule(l{2pt}r{2pt}){10-11} \cmidrule(l{2pt}r{2pt}){12-13} \cmidrule(l{2pt}r{2pt}){14-14}
			 \textit{Rockwell~Automation/ Allen-Bradley MicroLogix 1400} & 1831 & 37.97 & 68.00 & 68.00 & 0.00 & 0.00 & 100.00 & 0.00 & 9028 & 38.87 & 73436.3 & 38.65 & 8.13 \\
			 \textit{Omron CJ and CS PLC} & 1579 & 32.75 & 0.00 & 64.47 & 0.00 & 63.39 & 62.06 & 64.47 & 5219 & 22.47 & 45755.3 & 24.08 & 8.77 \\
			 \textit{Schneider~Electric BMX P34} & 785 & 16.28 & 0.00 & 100.00 & 0.00 & 100.00 & 100.00 & 100.00 & 4590 & 19.76 & 38983.9 & 20.52 & 8.49 \\
			 \textit{Siemens S7-300} & 439 & 9.10 & 0.00 & 0.00 & 0.00 & 100.00 & 100.00 & 79.05 & 4096 & 17.64 & 29636.7 & 15.60 & 7.24 \\
			 \textit{Mitsubishi Melsec Q} & 188 & 3.90 & 0.00 & 0.00 & 0.00 & 0.00 & 100.00 & 0.00 & 293 & 1.26 & 2186.3 & 1.15 & 7.46 \\
			 \cmidrule(l{2pt}r{2pt}){2-3} 
			 \cmidrule(l{2pt}r{2pt}){10-11} \cmidrule{12-13} 
			 \textbf{$\sum_{Total}$} & 4822 & 100.00 & &  &  &  &  &  & & 100.00 & 189998.5 & 100.000 &  \\
			\bottomrule
		\end{tabular}
	\end{table*}
	
	The number of devices,
	in absolute and relative terms throughout the set evaluated in this work,
	shows the \textit{Rockwell Automation/Allen-Bradley MicroLogix 1400} being the most common device in the analysis.
	Remarkably,
	the \textit{Omron CJ and CS PLC Series} have a significantly lower relative amount of \acp{cve} and of the weighted \ac{cvss} than of overall devices.
	That indicates a higher state of security compared to other devices.
	At the same time,
	the mean \ac{cvss} value per device is highest for the \textit{Omron CJ and CS PLC Series},
	indicating few but impactful vulnerabilities.
	In contrast,
	\textit{Schneider~Electric BMX P34} and \textit{Siemens S7-300} series have a significant higher distribution in \acp{cve} and \ac{cvss} than of overall devices,
	indicating more and more severe vulnerabilities on these devices.
	Regarding the security objectives,
	a majority of almost every device is vulnerable to \ac{dos} attacks.
	Apart from that,
	information disclosure is a frequent threat,
	followed by escalations of privilege.
	No device is susceptible to repudiation attacks, 
	spoofing only occurs in \textit{Rockwell Automation/Allen-Bradley MicroLogix 1400} devices.	

	\section{Discussion}
	\label{sec:discussion}
	This section presents the implications of the findings which are discussed in a qualitative fashion.
	After that,
	honeypots are introduced and related to the findings.

	\subsection{Honeypots}
	Honeypots are computer resources in a network that are solely intended to mimic real behaviour and thus attract attackers \cite{Zhang.2003}.
	They can be used in any environment.
	Consequently,
	honeypots tailored for \acp{ics} were designed,
	e.g. Conpot \cite{Conpot.2020}.
	\ac{ics} honeypots present an attacker the interface of an alleged industrial system,
	e.g. a water processing facility or a power plant~\cite{Navarro.2018}.
	The intention is twofold:
	First,
	insights about the attacker are gathered.
	The credentials and commands used by the attacker as well as indications of tools and goals can provide insight about the motives and aims of an attacker.
	Second,
	it binds resources of an attacker that consequently cannot be used to attack real systems,
	thus serving as a distraction.
	\textit{Shodan} provides a heuristic to detect honeypots,
	however,
	this was not applicable in this work.
	It is likely that some of the results are honeypots.
	The devices presented in Section~\ref{sec:evaluation} of the Mitsubishi devices which belong to a Japanese research institution are most likely research honeypots.
	Since these devices presented a valid fingerprint,
	they could have been real in the sense that it was hardware also used in productive environments,
	however,
	not connected to a process environment.
	These types of honeypots are solely distinguishable by analysing the correlation of actuator input and sensor output.
	However,
	this was the only instance of an obvious allocation of IP addresses by an individual institute that was encountered in the course of this work.

	\subsection{Discussion}
	The previous chapter showed that,
	first,
	there is a substantial number of \ac{ot} devices connected to the Internet and second,
	the majority of devices is susceptible to at least one known vulnerability.
	The implications are that either operators connecting their devices to the Internet against best practices do not employ best practices for security in general,
	or patching and updating \ac{ot} devices is too difficult to be feasible in praxis.
	Characteristics of \ac{ot} devices include spatial distribution over a large area,
	continuous operation with little to no time for maintenance and operating times of up to several decades,
	resulting in legacy systems.
	Due to these characteristics it is plausible that update- and patch-management is difficult in \ac{ot} environments.
	The absolute numbers,
	as well as consideration of past incidents show the threat to \ac{ot} environments.
	Furthermore,
	the effects range from low due to limited influence and knowledge of an attacker to significant damage to machines and products as well as threats to human life.
	All relevant attacker objectives of the STRIDE methodology can be achieved by \acp{cve}.
	The \ac{cvss} severity rating indicates how much impact on the network environment is to be expected in case of a successful attack.
	The power outage in the Ukraine or the destruction of nuclear processing plants in Iran demonstrate the destructive abilities.
	As mentioned in Section~\ref{sec:introduction},
	effects of cyber attacks on the \ac{ot} domain are not limited to the digital domain but can affect the physical domain as well.
	This property in combination with the large attack surface present a grave danger.
	For ethical and legal reasons it was not possible to try to exploit the vulnerabilities,
	so whether the identified vulnerabilities were exploitable or not was not evaluated.
	In general,
	there is no silver bullet for cyber security.
	Best practices need to be applied and even then,
	vulnerabilities can be found and exploited.
	A set of recommendations for \ac{ot} operators is,
	similar to home and office operators,
	to:	
	\begin{itemize}
		\item Use network segmentation. In case of \ac{ot} networks, they should be separated from \ac{it} networks and, especially, the public internet with firewalls and \acp{dmz}. Those are well-established, easy-to-implement solutions that go a long way.
		\item Deactivate services that are not used. This leads to a decreased attack surface.
		\item Activate security properties that are available in \acp{ics} to make spoofing more difficult once an attacker has gained access to a network.
		\item Implement \ac{iam} schemes, to restrict the attacker's capabilities.
	\end{itemize}
	
	Taking those recommendations into account will drastically reduce the threat.
	Additionally,
	vendors of \ac{ot} devices should adhere to security practices established in \ac{it} development.
	Patch management,
	security-by-design,
	and security features as a core concept should be implemented in devices by default. \par
	
	The geo-spatial analysis is intended to reveal insights about the importance of cyber security for \ac{ot} in certain regions.
	E.g. it could have been expected that richer countries like European or Northern American would put a higher focus on security.
	However, 
	most vulnerabilities of all devices were discovered in Europe and the USA,
	indicating that there are still challenges regarding \ac{ot} security.
	Furthermore,
	a correlation between legislation and the security state can be derived in a later work.

	\section{Conclusion}
	\label{sec:conclusion}
	This work intended to perform an \ac{osint}-based reconnaissance of vulnerabilities in \ac{ics} devices that could be exploited from the Internet.
	In total,
	more than \numprint{13000} specific devices or classes of devices were found of which virtually every device was potentially vulnerable to at least one \ac{cve}.
	Even though it required manual effort,
	the process of identifying devices is simple.
	It is important to note that the vendors producing the devices analysed in the course of this work are not exceptionally susceptible to cyber attacks.
	They were chosen because a significant amount of devices could be found to be analysed.
	It is expected that other vendors are fighting with the same security issues than the vendors analysed in this work.
	In the course of this work,
	the exploitability of the discovered vulnerabilities was not evaluated so that no conclusion about the concrete effects can be provided.
	Still,
	this allows for easy access to attackers with malicious intent.
	Furthermore,
	the potential effects of a successful exploitation can be derived from the attacker objectives achieved by the individual \acp{cve}.
	This provides a sound overview of the damage potential.
	However,
	so far,
	severe consequences were solely observed after explicit and thoughtful malicious activities,
	considered to be performed by professional groups of criminals,
	or state-sponsored actors.
	Furthermore,
	the attribution of \acp{cve} to the STRIDE features shows the potential effects an attack can have on an industrial environment.
	Generally,
	\ac{dos} conditions are the most frequently occurring attacks.
	As devices in production environments heavily rely on the availability of all entities,
	disrupting the availability can create a drastic effect on the process,
	in terms of loss of money as well as damage and spoilage of materials.
	Since the attribution of attacks to attackers is difficult~\cite{Fraunholz.2017d,Fraunholz.2017f},
	there is no way of being certain about the attackers,
	unless they are caught by law enforcement.
	It should be noted that the devices analysed are not specifically part of an \ac{iiot} environment.
	However,
	\ac{iiot} devices would employ the same infrastructure,
	thus,
	implementing strong cyber security solutions is a prerequisite to safe and secure \ac{iiot} environments.
	Still,
	the potential attack surface is significant,
	and the expected effects are severe.
	Industrial enterprises must put a stronger focus on security of \ac{ics} and \ac{ot} devices in order to prevent further successful attacks and consequent implications on real world domains.

	\section*{Acknowledgment}
	This work has been supported by the German Federal Ministry of Education and Research (BMBF) (Foerderkennzeichen 01IS18062E, SCRATCh).

	\ifCLASSOPTIONcaptionsoff
	\newpage
	\fi

	
	
	\bibliographystyle{IEEEtran}
	\bibliography{bibliography}
\end{document}